\documentclass[11pt, a4paper]{article}
\usepackage{jheparxiv}
\usepackage[utf8]{inputenc}
\usepackage{amsmath}
\usepackage{amsfonts}
\usepackage{amssymb}
\usepackage{latexsym}
\usepackage{mathrsfs}
\usepackage{graphicx}
\usepackage{color}
\usepackage{slashed}
\usepackage{twistor}
\usepackage[all]{xy}

 \newcommand{\rff}[1]{(\ref{#1})}

\renewcommand{\d}{\mathrm{d}}

\def \no {\nonumber}
\def \iffa {\iffalse} 

\def \ed  {\end{document}}

  \def \vp {\varphi}
  \def \a {\alpha}
  \def \b {\beta}

\def \ci {\cite} \def\ci{\cite}
\def\foot{\footnote}
\def \ed  {\end{document}}
 \def \td {\tilde}  

\def \ha {{\te {1\over 2}}}   
\def \ov {\over}
 \def \lab {\label} 
 \def \te {\textstyle} \def \b {\beta}
 \def \foot {\footnote} \def \ha {{\te{1\ov 2}}} 
\def \del {\partial}\def \vp {\varphi} \def \bb {{\rm b}} \def \fo {{\tfrac{1}{4}}} \def \z {\zeta} 
\def \A  {{\rm A}}  \def \B  {{\rm B}}
 \def \au {u}  \def \bu {v}

\begin{document}

\subheader{\hfill \texttt{Imperial-TP-TA-2018-01}}

\title{Scattering of conformal higher spin fields}

\author{Tim Adamo, Simon Nakach and Arkady A. Tseytlin\footnote{Also at Lebedev Institute, Moscow.}}

\affiliation{Theoretical Physics Group, Blackett Laboratory \\
        Imperial College London, SW7 2AZ, United Kingdom}

\emailAdd{[t.adamo,simon.nakach09,tseytlin]@imperial.ac.uk}

\abstract{We develop a formalism for describing the most general notion of tree-level scattering amplitudes in 4d conformal higher spin theory. As conformal higher spin fields obey higher-derivative equations of motion, there are many distinct on-shell external states which may contribute to their scattering, some of which grow polynomially  with time, leading to ill-defined amplitudes. We characterize the set of admissible  scattering states which produce finite tree amplitudes, noting that there are more such states than just standard massless higher spins obeying two-derivative equations of motion.  We use conformal gravity as a prime example, where the set of scattering states includes the usual Einstein graviton and a `ghost' massless spin 1 particle. An extension of the usual spinor helicity formalism allows us to encode these scattering states efficiently in terms of `twistor-spinors'. This  
leads to compact momentum space expressions for all finite tree-level 3-point amplitudes of conformal higher spin theory. While some of these 3-point amplitudes vanish (including all those with only  standard  two-derivative higher spin external states), there are  many others which are non-vanishing. We also comment on the generalization to scattering of conformal higher spins  in AdS$_4$. }

 
\maketitle

\setcounter{footnote}{0}
\section{Introduction}

Conformal higher spin (CHS)   theory  
(see, e.g.,  \ci{Fradkin:1985am,Fradkin:1989md,Fradkin:1990ps,Tseytlin:2002gz,Segal:2002gd,Metsaev:2007fq,Metsaev:2007rw,Bekaert:2010ky,Tseytlin:2013jya,Beccaria:2014jxa,Haehnel:2016mlb,Beccaria:2016syk,Adamo:2016ple,Grigoriev:2016bzl,Beccaria:2017nco}) 
is  a  formally  consistent 
 higher spin  model   that has a local action  with a flat-space vacuum  that generalizes spin $1$  Maxwell  theory and spin 2 
 Weyl gravity  to all  spins.\foot{In this paper we only consider the case of 4 dimensions and  bosonic integer spin  fields.} 
Locality  for a symmetric traceless   higher-spin field $\phi_s= ( \phi_{a_1...a_s})$ 
implies the presence  of $2s$ derivatives in the kinetic  term and thus  non-unitarity. Despite this, there are many reasons why CHS theory is an interesting topic for study, including its good UV behaviour,  relationship with other conformal field theories and higher-spins in anti-de Sitter space 
(cf.  \cite{Liu:1998bu,Tseytlin:2002gz,Metsaev:2009ym,Giombi:2013yva,Giombi:2014iua,Beccaria:2016tqy}).  

In such a higher-derivative theory, the definition of asymptotic states and scattering amplitudes is non-trivial. Given  the free  spin $s>1 $ CHS equation in transverse traceless  gauge  $\Box^s  \phi_s=0 $,  one can always choose a special  solution $\phi^{(0)}_s $  satisfying the 2-derivative     spin $s$  equation 
  $\Box  \phi^{(0)}_s=0 $.
 The latter   has  further on-shell gauge  invariance which reduces the number of independent  solutions 
   to the two   of the  standard   massless spin $s$  particle. 
   These  massless   spin $s$ degrees  of freedom  may be interpreted as  `physical'  (`unitary')  
  ones  while the rest of the  $s(s+1)$   degrees of freedom of a CHS field 
   are ghost-like.\foot{For example, in the case of the conformal gravity with $C^2$  action 
      this separation becomes obvious upon adding to the action an Einstein $R$ term or by 
      switching on a constant   curvature  resulting in   the ghost mode decoupling from the spin 2 Einstein  one 
   (and becoming massive in the first case  \ci{Stelle:1977ry}  and partially massless \ci{Deser:1983tm}   in the second).}
   
  Ignoring the fact that  the  `physical'   and `ghost'   degrees of freedom do not actually  separate on a flat   background 
  (in particular, in the sense   of  Hilbert space of asymptotic states  \ci{Berkovits:2004jj}),
  it is  natural to define  the CHS   scattering amplitudes  by keeping only  those  standard  massless   spin $s$ modes 
  on external lines  \ci{Beccaria:2016syk,Adamo:2016ple}. This is equivalent to defining the scattering amplitudes with the usual, two-derivative LSZ reduction.
  The underlying infinite-dimensional symmetry  algebra of the CHS theory   appears to imply that the
   corresponding tree-level S-matrix is trivial \ci{Joung:2015eny,Beccaria:2016syk,Adamo:2016ple}.

  One may wonder  if, at least at tree-level, a more general definition of the CHS S-matrix is possible. That is, can 
  the space of scattering states be extended to also include some of the `ghost'  modes
  leading to  non-vanishing   scattering amplitudes?
  This is the question we address in this paper. 
  As we shall  see,  in a momentum representation 
  one   can separate the  set of   `ghost'  modes into oscillating modes and modes whose curvature grows in time. The latter lead to formally infinite contributions 
   to  the on-shell action or   tree-level scattering amplitudes 
     and thus  appear  to be unsuitable  for scattering. However, the  oscillating modes can be  included, along with the massless  modes, into the set of admissible scattering states.

With this 
extended definition for the tree-level S-matrix, one still faces the obstacle of finding an efficient formalism for encoding the on-shell scattering states. Unlike a two-derivative theory in four dimensions, an on-shell CHS state is not uniquely specified by its momentum, spin, and positive/negative helicity label. Additional `polarization' data is required to distinguish how a spin $s$ state decomposes 
into the $s(s+1)$ degrees of freedom of CHS theory. 

We observe that this issue can be resolved by augmenting the standard spinor helicity formalism to include states carrying a spinor index of the conformal group. Such indices are also known as twistor indices. The resulting twistor-spinor formalism allows us to encode all on-shell states of CHS theory, and can be used to provide compact expressions which capture all finite tree-level 3-point amplitudes of the theory. This is analogous to the way in which massive states can be described by augmenting the spinor helicity formalism to include spinor 
indices of the little group~\cite{Conde:2016vxs,Conde:2016izb, Arkani-Hamed:2017jhn}.

\medskip

The plan of the paper is as follows: In section~\ref{Smatrix}, we give the generalised definition of the tree-level S-matrix in CHS theory and clarify which external states are admissible for scattering. We focus in detail on the example of conformal gravity, where in addition to the massless spin 2 Einstein graviton there is also an oscillating `ghost' spin 1 mode which can be used as a scattering state.\foot{The oscillating mode   corresponding to the  massless    spin 1  state  is  ghost-like as  it  originates from a time-like component of the metric fluctuation and thus contributes  with negative sign to the energy.}
We then describe the scattering states for a generic spin $s$ CHS field.\foot{Related work 
    describing possible  3-point vertices  for  higher derivative spin 0 and  1 conformal  fields in general
     dimension using a 2-derivative  formulation \ci{Metsaev:2007fq,Metsaev:2007rw} 
     (and thus   including  effectively all modes as  external states) 
     appeared in \ci{Metsaev:2016rpa}.}

In general, the separation of the field potential into `oscillating' and `growing' modes is gauge-dependent, even though the characterisation of scattering states is not. In section~\ref{Twistorspinor} we give a useful gauge-invariant description of scattering states in conformal gravity based on the corresponding curvatures. The resulting twistor-spinor formalism encodes all on-shell states of conformal gravity. We also comment on the relation between polarization tensors, curvatures, and the double-copy representation of conformal gravity in terms of a 4-derivative vector theory~\cite{Johansson:2017srf}.

Using the twistor-spinor formalism, we give an expression for all 3-point tree amplitudes (with complexified kinematics) in conformal gravity in section~\ref{CGAmps}. By evaluating this expression on specific on-shell states, we find that the only non-vanishing 3-point amplitudes in conformal gravity involve two spin 1 states and one Einstein graviton; these are essentially the same as in the Einstein-Maxwell theory. Section~\ref{CHSAmps} extends the twistor-spinor formalism to generic spin $s$ CHS fields, and we obtain an expression for all 3-point tree amplitudes of CHS theory.

In Section~\ref{AdS}, we consider the generalisation to CHS scattering in a background with a cosmological constant. We 
comment on the structure of on-shell states and 
evaluate our expressions for  3-point 
amplitudes 
for massless higher spin states in an AdS$_4$ background. Appendix~\ref{Helicity} describes  the helicity structure of linearised conformal gravity in the two gauges discussed in section~\ref{Smatrix}. Appendix~\ref{DOFcount} explains the counting of on-shell CHS degrees of freedom and scattering states in terms of curvatures. Appendix~\ref{PTder} contains a derivation of the 3-point CHS amplitudes from the formulation of CHS theory in twistor space~\cite{Haehnel:2016mlb,Adamo:2016ple}.



\section{Free higher-derivative fields  and  the  S-matrix}
\label{Smatrix}

External states in any scattering process are given by free field solutions of the 
equations of motion. A free bosonic 
  CHS gauge field  in 4 dimensions is a rank $s$ totally symmetric tensor $\phi_{a_1\cdots a_s}(x)=\phi_{a(s)}(x)$  
defined up to linearised gauge transformations
\be\label{lingt}
\delta \phi_{a(s)}=\partial_{(a_1} \epsilon_{a_2\cdots a_s)} +\eta_{(a_1 a_2} \alpha_{a_3 \cdots a_s)} \,,
\ee
which are  the spin $s$ generalisation of infinitesimal local diffeomorphisms (parametrized by $\epsilon_{a(s-1)}$) and conformal transformations (parametrized by $\alpha_{a(s-2)}$). The free field equation
can be written as $P^{a(s)\,b(s)}\phi_{b(s)}=0$, where $P^{a(s)\,b(s)}$ is a 
 transverse and traceless differential operator of order $2s$ which is totally symmetric in its indices 
 \cite{Fradkin:1985am}.  

Since spin $s$ CHS fields satisfy higher-derivative equations of motion, they contain many more on-shell degrees of freedom  (d.o.f.)  than ordinary two-derivative fields. Indeed, at spin $s$ there are $s(s+1)$ on-shell d.o.f..
Many of these $s(s+1)$ d.o.f. correspond to growing modes 
  with polynomial (rather than pure oscillatory) asymptotic behaviour.  
  
Two of the  d.o.f. 
 correspond to the standard two-derivative massless higher spin fields, which sit trivially inside the space 
 of solutions to the linearised CHS equations.   
These two-derivative solutions are \emph{not} the only on-shell states in CHS theory which are suitable for scattering in Minkowski space. There are other on-shell states which are pure oscillatory \emph{and} obey the higher-derivative equations of motion in a strict sense,
 satisfying the linear CHS equation of order $2s$ without solving some other equation which is of order 2 in derivatives. 


\subsection{Tree-level scattering in higher-derivative theories}

The notion   of scattering amplitudes in higher derivative theories is fraught with potential issues, 
but at tree-level an elementary definition   can be applied as long as there is an associated action functional. In two-derivative theories, tree-level scattering amplitudes can be defined by simply computing the multi-linear part of the classical action evaluated on a particular solution which is built (recursively) from superpositions of solutions to the free equations of motion with specified asymptotic behaviour~\cite{Arefeva:1974jv,Jevicki:1987ax}. For instance, a 3-point amplitude is obtained by extracting the coefficient of $\varepsilon_{1}\, \varepsilon_{2}\, \varepsilon_{3}$ from the classical action evaluated on
\be\label{3pteval}
\Phi^{[3]}(x)=\sum_{i=1}^{3}\varepsilon_{i}\,\phi_{i}(x)\,,
\ee
while a 4-point amplitude is the coefficient of $\varepsilon_{1}\cdots\varepsilon_{4}$ in the action evaluated on
\be\label{4pteval}
\Phi^{[4]}(x)=\sum_{i=1}^{4}\varepsilon_{i}\,\phi_{i}(x) +\int \d^{d}y\,\Delta(x,y)\,\left.\frac{\delta\cL_{\mathrm{int}}}{\delta\Phi}(y)\right|_{\Phi=\sum_{i=1}^{4}\varepsilon_{i}\phi_{i}}\,.
\ee
Here, the $\{\phi_{i}\}$ are free field solutions which can be expanded in a basis of plane waves, 
 $\Delta(x,y)$ is the propagator (i.e., the inverse of the kinetic operator in the action), and $\cL_{\mathrm{int}}$ is 
 the interaction part of   the Lagrangian.

For a higher-derivative theory, the same procedure can be used; the only 
subtlety is what free field solutions $\{\phi_i\}$ to use as the external states. As a toy example, consider a
 four-derivative scalar theory on a flat background
\be\label{fds1}
S= S_0 +   S_{\rm int} \ , \ \ \ \quad 
S_0= \ha  \int \d^{4}x\, (\Box\phi)^2 \ , \ \ \ \quad    S_{\rm int}=\tfrac{1}{6} \lambda \int \d^{4}x\,\phi^3\,.
\ee
$S_0$ is  conformally invariant with the scalar  assigned the 
conformal weight zero   and the coupling $\lambda$  has mass dimension 4.
  The  free equation of motion
\be\label{fds2}
\Box^{2}\phi=0\, 
\ee
admits a two-parameter $(\A,\B)$   family of solutions in terms of plane waves:
\be\label{fds3}
\phi(x)=\left(\A + \B\,n\cdot x\right)\e^{\im\, k\cdot x}\,,  \qquad \qquad  k^2=0 \ , 
\ee
where 
 $n^{a}$ is an arbitrary time-like vector (i.e.  $n \cdot k\not=0$, 
 e.g., $n^a= (1, 0, 0, 0)$). 
  The mode parametrized by $\B$  then grows linearly in time; 
  this is precisely the \emph{growing} mode expected for a generic theory with higher-derivative equations of motion (cf.   \cite{Pais:1950za}). 
  

The pure oscillatory  $\A$-modes 
 in \eqref{fds3}  are clearly suitable for scattering -- 
they are just the usual plane wave solutions to the two-derivative wave equation.
 The growing  $\B$-modes, on the other hand, lead to un-defined (or divergent) amplitudes even at tree-level. 
 Evaluated on three oscillatory external states, $S_{\rm int} $   gives  the expected  finite  
  3-point amplitude of a cubic scalar theory (we ignore the  overall numerical  factors)
\begin{equation*}
\cM_3 \sim  {\lambda}\,\delta^{(4)}\!\Big(\sum_{i=1}^{3}k_i\Big)\,.
\end{equation*}
However, if one of the external states is a growing mode, one finds from \rff{fds1}
\be\label{fds5}
\cM_3  \sim {\lambda}\int\d^{4}x\,n\cdot x\,\e^{\im\,(k_1+k_2+k_3)\cdot x}\,,
\ee
which is undefined. One can interpret this  `amplitude' in a purely distributional sense as
\be\label{fds6}
\cM_3  \sim -\im {\lambda}\,n\cdot\frac{\partial}{\partial K}\,\delta^{(4)}\!\Big(\sum_{i=1}^{3}k_i\Big)\,, \qquad\qquad  K^{a}:=(k_1+k_2+k_3)^{a}\,,
\ee
and similarly for the 3-point interactions involving two or three of the growing modes. 
But if one wishes to obtain finite tree-level amplitudes, supported on overall 4-momentum conservation, 
 it is clear that the growing modes must be excluded from the set of allowed external states.

This motivates a definition of  the 
S-matrix in a higher-derivative theory: 
tree-level scattering amplitudes are given by extracting the same multi-linear piece of the action as usual, with the added constraint that the free fields $\{\phi_i\}$ are solutions of the equations of motion which lead to \emph{finite}, \emph{momentum-conserving} amplitudes. The modes which this definition singles out as admissible external states will be referred to as the set of \emph{scattering states} of the theory. In the case of the conformal scalar theory \eqref{fds1}, the set of scattering states is precisely the ordinary plane waves.



\subsection{Linearized spectrum of  
  conformal gravity}
  \label{2p2}


Based on the example of the four-derivative conformal scalar, it is tempting to assume that the scattering states in a generic higher-derivative theory are simply the two-derivative zero-rest-mass 
fields of appropriate spin. Indeed, one might conclude that the space of spin $s$ scattering states in CHS theory is composed of only two-derivative massless spin $s$ free fields; this would mean that there are only two such modes for each spin, one each of helicity $\pm s$~\cite{Fronsdal:1978rb}. This, in turn, would indicate that the definition for the tree-level S-matrix is actually equivalent to that of two-derivative theories: the standard two-derivative LSZ reduction singles out all of the admissible scattering states on the external legs.

If this were the case, the tree-level S-matrix of CHS theory would be  rather trivial: 
there is strong evidence that all amplitudes of such two-derivative external states in CHS theory are zero~\cite{Maldacena:2011mk,Adamo:2013tja,Joung:2015eny,Beccaria:2016syk,Adamo:2016ple}.\foot{
For discussions of relations between the Einstein  and Weyl  actions   see also \ci{Metsaev:2007fq,Anastasiou:2016jix}.}
Fortunately, there are other scattering states in CHS theory; we now explain how to define them in a gauge-invariant manner.
 
The truncation of CHS theory to the spin 2 sector serves to capture all essential features of the problem of classifying higher-derivative, higher-spin scattering states. The corresponding non-linear theory 
  is \emph{conformal gravity}, a four-derivative theory of gravity governed by the action
\be\label{CGact1}
S[g]=\frac{1}{2\,\varepsilon^2}\int \d^{4}x\,\sqrt{|g|}\, C_{abcd}\,C^{abcd}\,,
\ee
where $\varepsilon$ is a dimensionless coupling constant and $C_{abcd}$ is the Weyl curvature tensor of the metric. The $s=2$ version of the linearised gauge freedom \eqref{lingt} on a flat background is
\be\label{CGgt1}
\delta h_{ab}= \partial_{a}\epsilon_{b}+\partial_{b}\epsilon_{a} + \eta_{ab}\,\alpha\,,
\ee
with the gauge parameters $\epsilon_{a}$  and  $\alpha$ encoding local diffeomorphisms and conformal transformations, respectively. 

Conformal invariance can always be used to fix a traceless gauge for linear perturbations, $h^{a}_{a}=0$, with the remaining gauge freedom
\be\label{CGgt2}
\delta h_{ab}=\partial_{a}\epsilon_{b}+\partial_{b}\epsilon_{a}-\frac{1}{2}\,\eta_{ab}\,\partial^{c}\epsilon_{c}\,.
\ee
To determine the modes of the theory, one must fix this gauge freedom  and find  the solutions of the linearised equations of motion.

\subsubsection{Conformal  gauge}

One such gauge-fixing is  `conformal' gauge~\cite{Riegert:1984hf}:
\be\label{CGcg1}
V_{c}:=\frac{1}{3}\partial^{a}\partial^{b}\partial_{c}\,h_{ab}-\Box\partial^{a}h_{ac}=0\,, \qquad  \qquad h^a_a =0 \ . 
\ee
It has the benefit of reducing the free equations of motion of  conformal gravity around a Minkowski background, which are  
\be \Box^{2}h_{ab}  + \partial_{a}  V_b +  \partial_{b}  V_b - \frac{1}{2} \eta_{ab}\partial^c V_c =0 \ , 
\label{linbach1} \ee
 to the spin-2 analogue  of \eqref{fds2}:
\be\label{CGcg2}
\Box^{2} h_{ab}=0\,.
\ee
It is easy to see that plane wave solutions to this equation are given by
\be\label{CGcg3}
h_{ab}=\left(\A_{ab}+\B_{ab}\,n\cdot x\right)\e^{\im\,k\cdot x}\,, \qquad \qquad k^2 =0 \  , 
\ee
where $n^a$ again is a time-like vector and 
$\A_{ab}$  and  $\B_{ab}$ are symmetric traceless constant tensors 
 related by four algebraic constraints   following from  \eqref{CGcg1}:
\be\label{CGcg4}
(n\cdot k)\,\B_{cb}\,k^{b}-\frac{\im}{4}\,k^{a}k^{b}\,\A_{ab}\,k_{c}=0\,.
\ee
These conditions do not completely fix the freedom in $\A_{ab}$, $\B_{ab}$; there are residual gauge transformations of the form
\begin{align}
\delta \A_{ab} &= \im\Big(\au_{a}k_{b}+\au_{b}k_{a}-\frac{1}{2}\eta_{ab}\,\au\cdot k\Big)+\bu_{a}n_{b}+\bu_{b}n_{a}-\frac{1}{2}\eta_{ab}\,\bu\cdot n\,, \label{CGcg5} \\
\delta \B_{ab} &=\im \Big( \bu_{a}k_{b}+\,\bu_{b}k_{a}-\frac{1}{2}\eta_{ab}\,\bu\cdot k\Big) \,, \nonumber 
\end{align}
where $\au_{a}$, $\bu_{a}$ are constant vectors.
Thus  there are   $2\times 9 -4 = 14$   independent parameters in 
 $\A_{ab}$, $\B_{ab}$, and 8  residual gauge transformations parametrized by $\au_{a}$, $\bu_{a}$. This implies 
  that there are six overall on-shell d.o.f., matching the  known  counting for conformal gravity \cite{Fradkin:1981jc,Lee:1982cp}. 
  
The residual gauge freedom can be used to distribute these six d.o.f.  in such a way that $\A_{ab}$ contains four  while $\B_{ab}$ contains two. It can be shown (see appendix~\ref{Helicity}) that the two d.o.f. in $\B_{ab}$ have helicity $\pm 2$, while the four 
  in $\A_{ab}$ have helicity $\pm1$ and $\pm 2$~\cite{Riegert:1984hf}. The helicity $\pm2$ modes in $\A_{ab}$ are precisely the Einstein gravitons, which form a consistent two-derivative sub-sector of the theory. 

At this point, one might na\"ively say that we have characterised the scattering states of conformal gravity: the modes encoded by $\A_{ab}$ are pure oscillatory and therefore suitable for scattering, while those in $\B_{ab}$ are growing and will not lead to well-defined amplitudes. But this is statement is premature: the decomposition of the modes into growing and oscillatory metric perturbations is not gauge invariant!

\subsubsection{Transverse gauge}

To see this, consider instead of the conformal gauge \eqref{CGcg1} the transverse  gauge
\be\label{CGtg1}
\partial^{a}h_{ab}=0\,,\qquad \qquad h^a_a=0 \ . 
\ee
 In this gauge the free equations of motion still take the form \eqref{CGcg2}, so 
  that $h_{ab}$ is again given by   \eqref{CGcg3}.
   However, instead of the constraints \eqref{CGcg4}, now  the symmetric traceless matrices $\A_{ab}$, $\B_{ab}$ must obey
\be\label{CGtg2}
k^{a}\,\B_{ab}=0\,, \qquad\qquad  \im\,k^{a}\,\A_{ab}+n^{a}\,\B_{ab}=0\,.
\ee
It is straightforward to show that 
there is a residual gauge freedom  parametrized by a single constant vector $\au^{a}$:
\begin{align}
 \delta \A_{ab} &=\im\Big(\au_{a}k_{b}+\au_{b}k_{a}-\frac{2}{5}\eta_{ab} \au\cdot k\Big)- 
 \frac{\im\, \au\cdot k}{5\, n\cdot k}\left(k_{a}n_{b}+k_{b}n_{a}\right)\,, \label{CGtg3} \\
 \delta \B_{ab} &= \frac{2\,\au\cdot k}{5\,n\cdot k}\,k_{a}k_{b}\,. \nonumber
\end{align}
Once again, this leaves  us   with six on-shell d.o.f.: $2\times 9 - 2\times 4  - 4=6$. 
In this case $\A_{ab}$ and $\B_{ab}$ each encode modes of helicity $\pm 2$, but 
the helicity $\pm 1$ modes are encoded by a linear combination of \emph{both} matrices, 
i.e.  the spin-1  modes  appear  to pick up a growing part  (see appendix~\ref{Helicity} for details). 

This seems to contradict what we found in the conformal gauge \rff{CGcg1}, where the spin-1 modes were purely oscillatory. The resolution of this apparent paradox 
 lies in the fact that we are trying to characterise the scattering states by looking at metric perturbations, which are not gauge invariant. Instead, we should look at the linearised curvature tensors associated with $h_{ab}$, where the decomposition into   modes is independent of the gauge choice.

It is clear that if a metric perturbation is purely oscillatory, then its associated (linearised) curvature tensor will also be
 purely  oscillatory. In conformal gauge, this means that the curvatures associated with the helicity $\pm 2$ Einstein  graviton 
modes \emph{and} the helicity $\pm 1$  modes are purely oscillatory
while  the curvature associated with the helicity $\pm 2$ modes encoded by $\B_{ab}$ contains  linearly growing 
terms.
In the  transverse gauge, the Einstein modes have the same pure oscillatory curvatures, and the spin-2 modes encoded by $\B_{ab}$ have the same growing curvatures. The spin-1 modes are encoded by a metric perturbation of the form (cf. appendix~\ref{Helicity})
\be\label{CGtg4}
h_{ab}\sim (1-2\im\, n\cdot x)S_{ab} + (1+2\im\, n\cdot x)\tilde{S}_{ab}\,,
\ee
where $S_{ab}$, $\tilde{S}_{ab}$ are constant matrices whose entries are determined by $n_{a}$ and $k_{a}$. \emph{A priori}, the curvature associated with \eqref{CGtg4} could have a linearly growing piece, but an explicit calculation shows that these growing terms cancel. The remaining oscillatory curvature of course matches what was found in conformal gauge.

\subsubsection{Two-derivative formulation of conformal gravity}

To get another perspective on the spectrum of states in conformal gravity, it is  useful to consider  its 2-derivative reformulation   by introducing extra   fields  in addition to the metric. 
This  is a generalization  of replacing the 4-derivative scalar Lagrangian $\cL= \ha  (\Box \phi)^2 + V(\phi)$  by an  equivalent one 
with two independent fields: 
$\cL'= \vp  \Box \phi - \ha \vp^2 +    V(\phi) $.  The equations following from $\cL'$ are $\Box \phi=\vp$ and $\Box \vp=0$ so that 
$\vp$   is an  oscillating mode  while $\phi$ contains both oscillating and growing modes, with the scale of the 
latter  being related to $\vp$. Thus from the  point of view  of the original $\Box^2 \phi=0$   theory 
the field $\vp$ represents the growing mode  and should not be  included in the set of asymptotic states. 

Observing that in 4 dimensions  the Weyl Lagrangian  $C^2$   in \rff{CGact1} 
is the  same (up to a  total derivative)  as 
\be\lab{11}   \cL_W=\te  2 \sqrt{|g|} \Big(R^2_{ab} - {1\ov 3} R^2\Big) \ ,  \ee
 one can  introduce an auxiliary 
 tensor $\vp_{ab}$  to rewrite it in the $\vp R - \vp^2$  form. 
 Alternatively, one may start with a formulation  of Weyl  gravity as a gauge theory of the 
 SO$(2,4)$  conformal group \ci{Kaku:1977pa}
ending up with    \ci{Metsaev:2007fq} 
\be \lab{2}
\cL'_W= - \sqrt{|g|} \Big[  \vp^{ab} \hat G_{ab}  +  \fo  ( \vp^{ab} \vp_{ab} - \vp^a_a \vp^b_b) +   \fo  F^{ab} F_{ab} \Big]\ .
\ee
Here  $\vp_{ab}$ is related to the gauge  potential  corresponding to the
conformal boosts,   $F_{ab}= \del_a \bb_b - \del_b \bb_a$  
is  the  field strength of the gauge potential  $\bb_a$  of dilatations  and 
\be \lab{3}
 \hat G_{ab} \equiv  R_{ab} - \ha  g_{ab} R   +  \nabla_{(a} \bb_{b)}  + \ha \bb_a \bb_b  - g_{ab} ( \nabla^c \bb_c - \fo   \bb^c \bb_c) \ , 
 \ee
 where $\nabla_a$  is covariant derivative corresponding to $g_{ab}$. 
Integrating  out $\vp_{ab}$  in \rff{2} one finds  (ignoring total derivatives)  
 that   all dependence on $\bb_a$   cancels out 
  and one recovers the Weyl Lagrangian \rff{11}. 
  
The reason for the decoupling of  $\bb_a$ is that  the action  for  \rff{2} is invariant, 
in addition  to the usual reparametrizations and Weyl rescalings of the metric $\delta g_{ab} = \lambda  g_{ab}$, under 
\be \lab{4}
\delta \vp_{ab} =  2\nabla_{(a} \z_{b)}  +  2\bb_{(a} \z_{b)}    - g_{ab}   \bb^c \z_c \ , \qquad \qquad 
\delta \bb_a = \del_a \lambda  - \z_a \ . \ee
Here $\z_a$   is the parameter of gauge transformations  corresponding to local conformal boosts.
One can   fix this   symmetry   by setting $\bb_a=0$, but  instead 
one  may impose a gauge   condition on $\vp_{ab}$.

  The linearised  
equations of motion for the fields  $h_{ab}= g_{ab}-\eta_{ab}$,  $\vp_{ab}$ and $\bb_a$   that follow from  \rff{2} are 
\begin{align}  \lab{5}
&\vp_{ab} = -2 ( \hat G_{ab} - \tfrac{1}{3}   g_{ab} \hat G^c_c) = - 2 (R_{ab} - \tfrac{1}{6} g_{ab} R)  -   2 \del_{(a} \bb_{b)} +     ... \ , \\ 
&\Delta_2 \vp_{ab} + ...=0  \ , \lab{6}\\
&   \del^a  F_{ab} +  \del^a \vp_{ab}  - \del_b\vp^c_c  + ...=0 \ . \lab{66}
\end{align} 
Here $R_{ab} = - \ha  \Box h_{ab}   + \del_{(a} \chi_{b)} + O(h^2), \ \chi_a = \del^b  h_{ab}  - \ha  \del_a h$  and 
 we defined  $\Delta_2$  as the linearized Einstein operator:
$R_{ab} - \ha  g_{ab}  R =  \ha \Delta_2 h_{ab}+ O(h^2)$,  $\Delta_2 =- \Box  + \cdots$. 
Fixing the  reparametrizations and Weyl rescalings by 
 the TT gauge \rff{CGtg1} on   $h_{ab}$   and  fixing the $\z_a$-symmetry  \rff{4} by  the  harmonic gauge on $\vp_{ab}$ 
 \be \lab{7}
 \del^a  h_{ab} =0 \ , \ \ \    h^a_a=0 \ ; \ \ \ \ \ \ \ \ \   
 \del^a \vp_{ab}=\ha \del_b  \vp^c_c  \ , \ee
it follows from  eqs.\rff{5},\rff{6} (and their traces and derivatives)  that 
 \be \lab{8}
\Box h_{ab} =  \vp_{ab} + 2 \del_{(a} \bb_{b)} \ ; \qquad  \Box \vp_{ab}=0 \ , \qquad 
\vp^a_a + 2  \del^a b_a =0 \ ; \qquad \Box b_a =0 \ , 
\ee
with  \rff{66} satisfied automatically. 
Using the  on-shell  gauge invariance of $\vp_{ab}$ we may then set $\vp^a_a=0$  and thus   $\del^a b_a =0$.

We then conclude   
 that $\vp_{ab}$ describes a massless  purely-oscillating graviton mode,  $\bb_a$ describes a  massless vector,   and 
 $h_{ab}$  contains the Einstein  graviton part  plus a  growing mode as in \rff{CGcg3} 
      with the  polarization tensor B$_{ab}$   of the  latter   being  related as in  \rff{8}   to the polarization tensors  of 
 $\vp_{ab}$   and $\bb_a$  via  the $\zeta$-gauge invariant combination $\vp_{ab} + 2 \del_{(a} \bb_{b)}$  (cf. \rff{4}).
This way we    recover the 2+2+2=6  count of on-shell degrees of freedom, with the scattering states being represented by 
the Einstein  graviton A$_{ab}$-part of $h_{ab}$ in \rff{CGcg3}  and  by  the  massless vector $\bb_a$.


An  advantage of the 2-derivative representation \rff{2}  in terms of the  3  fields $h_{ab}, \vp_{ab}$ and $\bb_a$ 
 is that  it formally extends  off-shell and  also to  the non-linear level. 
 In particular, it is implied by 
 the structure of  \rff{2}   that the 
 3-point scattering  of  one  Einstein  graviton   contained 
  in  $h_{ab}$  (A-mode) and   two  massless   vectors described by $\bb_a$  should 
 be  the same as in Einstein-Maxwell  theory, since such a vertex  
 may only come from the $F_{ab} F_{ac}  h_{bc}$   term  in \rff{2}. 
 We shall   reach the same  conclusion via  a different route  in section 4 below.
 It would  be interesting also to  use the 2-derivative action  \rff{2}  to compute the  4-point amplitudes 
 involving the vector $\bb_a$.

\subsection{Scattering  states in CHS theory}

To summarize the above discussion,  
the gauge-invariant characterisation of scattering states is in terms of their curvatures. Those modes which, in a plane wave basis, lead to purely oscillatory linearised curvature tensors are the suitable states for scattering. This is true even when, in a particular gauge, their potentials are growing. The space of scattering states in conformal gravity therefore contains  two helicity $\pm 2$  states  (Einstein graviton)  as well as two  helicity $\pm 1$ states (spin-1 mode).

From now on, we refer to a mode as `growing' (`oscillatory') if the mode's curvature is growing (oscillatory) in a plane wave basis. For CHS fields with $s>2$, the number of scattering states (or purely oscillatory modes) increases with $s$. The $s(s+1)$ on-shell d.o.f. for a free spin-$s$ CHS field~\cite{Fradkin:1985am}  can be 
divided equally into positive and negative helicity. Let $\nu_{s,h}$ be the number of on-shell d.o.f. in a negative helicity spin-$s$ CHS field which correspond to helicity $-h$, where $h=1,\ldots,s$. For $s=2$, we saw that $\nu_{2,1}=1$, $\nu_{2,2}=1+1=2$. In general, one can show that
\be\label{CHSdof1}
\nu_{s,h}=h\,,
\ee
which is consistent with the overall degree of freedom counting in the negative helicity sector: 
\be\label{CHSdof2}
\sum_{h=1}^{s}\nu_{s,h}=\tfrac{1}{2}\,  s(s+1) \ . 
\ee
Since the equations of motion are of order $2s$ in derivatives, many of these states will correspond to modes which grow at least linearly and at most of order $(s-1)$ in time. The number of such modes increases quadratically with $s$, so that for $s\geq3$ there are more growing modes than oscillatory modes. Writing $\nu_{s,h}=\hat{\nu}_{s,h}+\nu^{\circ}_{s,h}$, where $\hat{\nu}_{s,h}$ is the number of growing modes for the spin $s$ CHS field of helicity $-h$, it can be shown that (see appendix~\ref{DOFcount})
\be\label{CHSdof3}
\hat{\nu}_{s,h}=h-1 \quad \Rightarrow \quad \hat{\nu}_{s}:=\sum_{h=1}^{s}\hat{\nu}_{s,h}=\tfrac{1}{2}\,  s(s-1)   \,,
\ee
where $\hat{\nu}_{s}$ is the total number of growing d.o.f. at spin $s$. This indicates that the spin $s$ field contains $\nu^{\circ}_{s}=s$ purely oscillatory modes, and it follows that these are distributed such that there is a single one at each integer helicity $-1,\ldots,-s$. This decomposition is derived in appendix~\ref{DOFcount} from the structure
 of linearised field strengths. It  is also consistent with the  2-derivative formulation  of CHS fields
  \ci{Metsaev:2007fq,Metsaev:2007rw}\foot{Let us note  
 that a 2-derivative formulation 
 is known   for all CHS    fields  but  so far only at the quadratic level 
  \ci{Metsaev:2007rw}. The  existence of  such  a 2-derivative 
   local  action at an  interacting level   is  an open question   for $s>2$. One may  still  attempt to construct 
   interacting 3-point vertices using an indirect light-cone approach as  developed  for  low  $s<2$    spins in any dimension in \ci{Metsaev:2016rpa}.
  }    and  with the  structure of the free    CHS 
partition function  \ci{Tseytlin:2013jya}.

In summary, the set of spin-$s$ scattering states in CHS theory contains $2s$ different kinds of modes: one each of helicity $\pm1, \pm2,\ldots,\pm s$. The helicity $\pm s$ states are precisely the two-derivative massless
 higher spin fields; the others arise as oscillatory solutions of the higher-derivative CHS equations of motion.


\section{Twistor-spinor representation of   states of  conformal gravity}
\label{Twistorspinor}

Having established that the standard two-derivative LSZ-type  reduction misses out 
admissible asymptotic scattering states in a higher-derivative theory, a natural question is: how can the various different modes of a higher-derivative field be distinguished at the level of the free field? In this section, we develop a formalism that allows us to isolate the 
 on-shell modes of conformal gravity, which serves as a toy example of conformal higher spin theory. Indeed, the formalism extends naturally to CHS fields of all integer spins.

The basic idea is that CHS fields are most naturally represented in terms of objects which carry \emph{twistor} indices, as well as the usual spinor indices familiar from the spinor helicity formalism. These objects are simply tensors upon which a conformally invariant connection acts covariantly.\footnote{This connection is alternatively known as the local twistor connection or Cartan conformal connection on space-time~\cite{Cartan:1923,Dighton:1974,Friedrich:1977}.} After demonstrating how the free field equations of conformal gravity are recovered in this `twistor-spinor' formalism, we show how the d.o.f. decomposition is achieved for momentum eigenstates.


\subsection{Twistor-spinors and  linearised Bach equations}

\iffa
In terms of a linear metric fluctuation $h_{ab}$ (assumed to be trace-free), the free field equations of conformal gravity around a Minkowski background are:
\be \Box^{2}h_{ab}  + \partial_{a}  V_b +  \partial_{b}  V_b - \frac{1}{2} \eta_{ab}\partial^c V_c =0 \ , 
\label{linbach1} \ee
where $V_a$ was defined in \rff{CGgt1}. 
\begin{multline}\label{linbach1}
\Box^{2}h_{ab}+\frac{1}{3}\partial_{a}\left(\partial_{b}\partial^{c}\partial^{d} h_{cd}-3\Box\partial^{c}h_{bd}\right)+\frac{1}{3}\partial_{b}\left(\partial_{a}\partial^{c}\partial^{d} h_{cd}
-3\Box\partial^{c}h_{ad}\right) \\
-\frac{1}{6}\eta_{ab}\partial^{e}\left(\partial_{e}\partial^{d}\partial^{d}h_{cd}-3\Box\partial^{c}h_{ec}\right)=0\,.
\end{multline}
\fi

The   free field equations of conformal gravity  written in terms  of the metric fluctuation $h_{ab}$
in   \rff{linbach1}   can also be written in terms of the 
  linearised Weyl tensor  $C_{abcd}$ of $h_{ab}$   as    
\be\label{linbach2}
\partial^{a}\partial^{d}\,C_{abcd}=0\,,
\ee
which are often known as the linearised Bach equation. 

In four dimensions, the Weyl tensor decomposes into the self-dual (SD) and anti-self-dual (ASD) parts, given by totally symmetric spinors $\widetilde{\Psi}_{\dot\alpha\dot\beta\dot\gamma\dot\delta}$ and $\Psi_{\alpha\beta\gamma\delta}$, respectively. At the linear level, this is equivalent to the statement that the fluctuation $h_{ab}$ can be decomposed into positive and negative helicity parts. The free field equations for these helicity sectors are then
\be\label{linbach3}
\partial^{\alpha\dot\alpha}\,\partial^{\beta\dot\beta}\,\widetilde{\Psi}_{\dot\alpha\dot\beta\dot\gamma\dot\delta}=0\,, \qquad \partial^{\alpha\dot\alpha}\,\partial^{\beta\dot\beta}\,\Psi_{\alpha\beta\gamma\delta}=0\,,
\ee
with $\widetilde{\Psi}_{\dot\alpha\dot\beta\dot\gamma\dot\delta}$ corresponding to a positive helicity perturbation and $\Psi_{\alpha\beta\gamma\delta}$ a negative helicity perturbation. Spinor indices are raised and lowered using the 2d Levi-Civita symbols $\epsilon_{\alpha\beta}$, $\epsilon_{\dot\alpha\dot\beta}$, etc.

Solutions to the  standard zero-rest-mass equations of linearised Einstein gravity,
\be\label{linEin}
\partial^{\alpha\dot\alpha}\,\widetilde{\Psi}_{\dot\alpha\dot\beta\dot\gamma\dot\delta}=0\,, \qquad \partial^{\alpha\dot\alpha}\,\Psi_{\alpha\beta\gamma\delta}=0\,,
\ee
are trivially solutions to the four-derivative Bach equations \eqref{linbach3}, so Einstein gravitons form a subsector of the solutions. But of course there are other solutions to the Bach equations which are not strictly two-derivative in nature.

\medskip

Let us introduce a new kind of index, called a \emph{twistor index}, which can be carried by space-time fields (cf.   \cite{Penrose:1972ia,Penrose:1986ca}).\footnote{More precisely, this will be a \emph{local twistor index}, corresponding to a field valued in a rank four vector bundle over space-time whose fibres are copies of the (flat) twistor space of Minkowski space-time. We drop the `local' prefix for much of this paper, as we are not concerned with comparison to global twistors.} Twistor indices are equivalent to $\SL(4,\C)$ spinor indices, and will be denoted by $A,B,C,\ldots$; the four values of the index are decomposed into 2-spinors of opposite chirality. For instance, rank one covariant and contravariant twistor fields are decomposed into spinor fields as:
\be\label{ltf1}
T_{A}(x)=\left(\begin{array}{c}
                \tilde{t}_{\dot\alpha}(x) \\
                t^{\alpha}(x)
               \end{array}\right)\,, \qquad S^{A}(x)=\left(\begin{array}{c}
                                                            \tilde{s}^{\dot\alpha}(x) \\
                                                            s_{\alpha}(x)
                                                            \end{array}\right)\,.
\ee
Twistor indices can be paired with ordinary spinor indices, with the constraint that components of the resulting field are trace-free, for instance,  
\begin{equation}
 T_{A\beta}=\left(\begin{array}{c}
                   \tilde{t}_{\dot\alpha \beta} \\
                   t^{\alpha}{}_{\beta}
                  \end{array}\right)\,,  \qquad \qquad t^{\alpha}{}_{\alpha}=0 \ . 
\end{equation}
The key property of twistor indices is that they are acted on by a particular conformally-invariant connection on space-time, known as the (local) twistor connection or Cartan conformal connection. On any 4d space-time, this twistor connection is given locally by 
\be\label{1}
 \D_{\alpha\dot\alpha}=\nabla_{\alpha\dot\alpha} +\CA_{\alpha\dot\alpha} \ , \ee
 where $\nabla_{\alpha\dot\alpha}$ is the Levi-Civita connection and the 1-form $\CA$ takes values in the (complexified) conformal algebra $\mathfrak{sl}(4,\C)$. In terms of geometric data, the potential $\CA$ is
\be\label{LTcon1}
(\CA_{\alpha\dot\alpha})^{B}{}_{C}=\left(\begin{array}{c c}
                                          0 & \delta_{\alpha}^{\gamma}\,\delta_{\dot\alpha}^{\dot\beta} \\
                                          -P_{\alpha\dot\alpha\beta\dot\gamma} & 0
                                         \end{array}\right)\,,
\ee
where $P_{\alpha\dot\alpha\gamma\dot\beta}$ is the Schouten tensor,
\be\label{schouten}
P_{\alpha\dot\alpha\beta\dot\gamma}:=\Phi_{\alpha\beta\dot\alpha\dot\gamma}-\Lambda\,\epsilon_{\alpha\beta}\,\epsilon_{\dot\alpha\dot\gamma}\,,
\ee
written in terms of the trace-free Ricci curvature $\Phi_{\alpha\beta\dot\alpha\dot\gamma}$ and scalar curvature $\Lambda$ of the Levi-Civita connection. The action of this connection on twistor-indexed quantities is given by the rule:
\be\label{LTcon2}
\D_{\alpha\dot\alpha} T^{B}=\nabla_{\alpha\dot\alpha} T^{B}+(\CA_{\alpha\dot\alpha})^{B}{}_{C}\,T^{C}\,, \qquad \D_{\alpha\dot\alpha}S_{B}=\nabla_{\alpha\dot\alpha} S_{B}+(\CA_{\alpha\dot\alpha})_{B}{}^{C}\,S_{C}\,,
\ee
and the action on higher valence twistor indices can be deduced by the Leibniz rule (cf.   \cite{Penrose:1986ca}).

\medskip

The reasons for considering twistor-valued objects on space-time are two-fold. First of all,   objects with  twistor indices are spinors of the conformal group and the twistor connection is itself conformally invariant. This is evident from the curvature of the connection:
\be\label{LTcurv}
[\D_{\alpha\dot\alpha},\,\D_{\beta\dot\beta}]=(F_{\alpha\dot\alpha\beta\dot\beta})^{C}{}_{D}=\left(\begin{array}{c c}
                                                                                                    \epsilon_{\alpha\beta}\,\widetilde{\Psi}_{\dot\alpha\dot\beta\dot\delta}{}^{\dot\gamma} & 0 \\
                                                                                                    (\epsilon_{\beta\alpha}\nabla_{\gamma}{}^{\dot\rho}\widetilde{\Psi}_{\dot\alpha\dot\beta\dot\delta\dot\rho} + \epsilon_{\dot\beta\dot\alpha}\nabla_{\dot\delta}{}^{\rho} \Psi_{\alpha\beta\gamma\rho}) & \epsilon_{\dot\beta\dot\alpha}\,\Psi_{\alpha\beta\gamma}{}^{\delta}
                                                                                                   \end{array}\right)\,,
\ee
which is conformally covariant. Hence, this formalism is ideally suited to describing a conformally invariant theory such as conformal gravity. The second reason is that first-order equations of motion with respect to the local twistor connection often correspond to higher-derivative equations on the components of a twistor-valued object~\cite{Mason:1987,Mason:1990}.

\medskip

Consider a twistor-spinor field of the form
\be\label{lcg1}
\Gamma_{A\beta\gamma\delta}=\left(\begin{array}{c}
                                   \gamma_{\dot\alpha\beta\gamma\delta} \\
                                   \Psi^{\alpha}{}_{\beta\gamma\delta}
                                  \end{array}\right)\,,
\ee
where $\Gamma_{A\beta\gamma\delta}=\Gamma_{A(\beta\gamma\delta)}$ and the component $\Psi^{\alpha}{}_{\beta\gamma\delta}$ obeys
\be\label{lcg2}
\Psi^{\alpha}{}_{\alpha\gamma\delta}=0 \quad \Rightarrow \quad \Psi_{\alpha\beta\gamma\delta}=\Psi_{(\alpha\beta\gamma\delta)}\,.
\ee
Let us treat $\Gamma_{A\beta\gamma\delta}$ as a linear field on Minkowski space-time, and impose a free equation of motion using the twistor connection in \rff{1}:
\be\label{lcg3}
\D^{\beta\dot\beta}\,\Gamma_{A\beta\gamma\delta}=0\,.
\ee
From \eqref{LTcon1} and \eqref{LTcon2}  this  is equivalent to a system of equations for the components of $\Gamma_{A\beta\gamma\delta}$:
\be\label{lcg4}
\partial^{\beta\dot\beta}\,\gamma_{\dot\alpha\beta\gamma\delta}=0\,, \qquad \partial^{\beta\dot\beta}\,\Psi^{\alpha}{}_{\beta\gamma\delta}-\gamma^{\dot\beta\alpha}{}_{\gamma\delta}=0\,.
\ee
The second of these equations defines $\gamma_{\dot\alpha\beta\gamma\delta}$ in terms of $\Psi_{\alpha\beta\gamma\delta}$ on-shell, which can be substituted back into the first equation to give
\be\label{lcg5}
\partial^{\alpha\dot\alpha}\,\partial^{\beta\dot\beta}\,\Psi_{\alpha\beta\gamma\delta}=0\,,
\ee
i.e. the negative helicity free-field equation of conformal gravity in  \eqref{linbach3}.

A similar story holds for the positive helicity free field equation. In this case, one has a twistor-spinor field
\be\label{lcg6}
\widetilde{\Gamma}^{A}{}_{\dot\beta\dot\gamma\dot\delta}=\left(\begin{array}{c}
                                                                \widetilde{\Psi}^{\dot\alpha}{}_{\dot\beta\dot\gamma\dot\delta} \\
                                                                \tilde{\gamma}_{\alpha\dot\beta\dot\gamma\dot\delta}
                                                               \end{array}\right)\,, \qquad \widetilde{\Psi}^{\dot\alpha}{}_{\dot\alpha\dot\gamma\dot\delta}=0\,,
\ee
obeying an equation of motion
\be\label{lcg7}
\D^{\beta\dot\beta}\,\widetilde{\Gamma}^{A}{}_{\dot\beta\dot\gamma\dot\delta}=\left(\begin{array}{c}
                                                                                     \partial^{\beta\dot\beta}\widetilde{\Psi}^{\dot\alpha}{}_{\dot\beta\dot\gamma\dot\delta}-\tilde{\gamma}^{\beta\dot\alpha}{}_{\dot\gamma\dot\delta} \\
                                                                                     \partial^{\beta\dot\beta}\tilde{\gamma}_{\alpha\dot\beta\dot\gamma\dot\delta}
                                                                                    \end{array}\right) =0\,,
\ee
in Minkowski space. Just like the negative helicity case, the coupled equations \eqref{lcg7} imply that $\widetilde{\Psi}_{\dot\alpha\dot\beta\dot\gamma\dot\delta}$ obeys the positive helicity free field equation of conformal gravity.

It should be noted that there is also a description of \emph{non-linear} conformal gravity in terms of the twistor connection. Indeed, conformal gravity is equivalent, at the non-linear level, to a gauge theory of the twistor connection, with the Bach equations given by the Yang-Mills equations of the twistor connection~\cite{Merkulov:1984nz}.


\subsection{Momentum eigenstates}

To see the counting of on-shell states, it is useful to go to a momentum eigenstate basis to find solutions of \eqref{lcg3} and \eqref{lcg7}. Consider a negative helicity field constructed as:
\be\label{nh1}
\Gamma_{A\beta\gamma\delta}=B_{A}\,\lambda_{\beta}\,\lambda_{\gamma}\,\lambda_{\delta}\,\e^{\im\,k\cdot x}\,,
\ee
where $k_{\alpha\dot\alpha}=\lambda_{\alpha}\tilde{\lambda}_{\dot\alpha}$ is an on-shell (massless) 4-momentum.\footnote{Our conventions for the 4-dimensional spinor helicity formalism follow~\cite{Adamo:2017qyl}. Dotted $\SL(2,\C)$ spinor indices are positive chirality, un-dotted $\SL(2,\C)$ spinor indices are negative chirality and we use the notation: $[\tilde{\lambda}\,\tilde{\beta}] \equiv  \tilde{\lambda}^{\dot \b} \,\tilde{\beta}_{\dot \b} =\tilde{\lambda}^{\dot \b} \,\tilde{\beta}^{\dot \a}  \epsilon_{\dot \a \dot \b}  $, \ \ 
$\la \beta \lambda \ra \equiv   \beta^\b \lambda_\b =  \b^\b  {\lambda}^\a  \,   \epsilon_{\a\b}$, etc.} The operator $B_{A}=(\tilde{B}_{\dot\alpha},\,B^{\alpha})$ is a \emph{helicity lowering operator}, since its role is to convert the helicity $-\frac{3}{2}$ Rarita-Schwinger field into a negative helicity conformal gravity field (which must include helicity $-2$ fields). Its components have mass dimension 
\be\label{nh2}
[\tilde{B}_{\dot\alpha}]=\frac{1}{2}\,, \qquad [B^{\alpha}]=-\frac{1}{2}\,,
\ee
as dictated by conformal invariance.

The components of this operator are constrained by a single condition, which descends from the twistor geometry underlying the construction (cf.   \cite{Berkovits:2004jj,Adamo:2013tja}). Define the following twistor-indexed differential operator, which acts on on-shell momenta:
\be\label{dop}
C^{A}:=\left(-\im \frac{\partial}{\partial\tilde{\lambda}_{\dot\alpha}},\,\lambda_{\alpha}\right)\,.
\ee
The condition imposed on $\Gamma_{A\beta\gamma\delta}$ is
\be\label{nh3}
C^{A}\,\Gamma_{A\beta\gamma\delta}=0\,.
\ee
Assuming that the components of $B_{A}$ obey
\be\label{nh4}
\frac{\partial B^{\alpha}}{\partial\tilde{\lambda}_{\dot\beta}}=0\,, \qquad \frac{\partial}{\partial x^{\beta\dot\beta}}\tilde{\lambda}^{\dot\alpha}\tilde{B}_{\dot\alpha}=0\,,
\ee
the constraint \eqref{nh3} becomes a simple PDE in on-shell momentum space:
\be\label{nh5}
\frac{\partial\tilde{B}_{\dot\alpha}}{\partial\tilde{\lambda}_{\dot\alpha}}+\im\,\lambda_{\alpha}\,B^{\alpha}=0\,.
\ee
This can be solved for $\tilde{B}_{\dot\alpha}$ to give:
\be\label{nh6}
\tilde{B}_{\dot\alpha}=\frac{\partial B}{\partial\tilde{\lambda}^{\dot\alpha}} - \,\frac{\im }{2}\, \tilde{\lambda}_{\dot\alpha}\lambda_{\alpha}\,B^{\alpha}\,,
\ee
which reduces the d.o.f. to three, parametrized by $\{B,\,B^{\alpha}\}$, 
matching  the  on-shell counting for conformal gravity  discussed above.

Since there are three total d.o.f., we should be able to construct three distinct states by making different choices for $\{B,\,B^{\alpha}\}$. These choices are, of course, constrained by the fact that the resulting fields must satisfy the equations of motion. To begin, consider a negative helicity Einstein graviton: this is a field whose Weyl spinor $\Psi_{\alpha\beta\gamma\delta}$ trivially satisfies \eqref{lcg5} by virtue of obeying the zero-rest-mass equation $\partial^{\alpha\dot\alpha}\Psi_{\alpha\beta\gamma\delta}=0$.

By singling out this two-derivative solution inside the space of solutions to the four-derivative Bach equations, mass scales are introduced by necessity. One of these is the coupling constant of Einstein gravity $\kappa= \sqrt{8\pi G_{\mathrm{N}}}$: the coupling $\varepsilon$ of conformal gravity in \rff{CGact1}  is dimensionless, whereas $\kappa$ has mass dimension $-1$. Furthermore, conformal gravity cannot distinguish between Minkowski space and (A)dS$_4$, since these backgrounds are related by a conformal transformation. Selecting the Einstein solution therefore introduces a cosmological constant $\Lambda$, of mass dimension $+2$. Since we consider a Minkowski background, this corresponds to $\Lambda=0$.

With this in mind, a negative helicity Einstein graviton corresponds to $B^{\alpha}\propto\lambda^{\alpha}$; it remains to determine the constant of proportionality. Since $B^{\alpha}$ has mass dimension $-\frac{1}{2}$, the dimensionful scales of this solution allow us to fix $B^{\alpha}=\kappa\,\lambda^{\alpha}$. The equations of motion set $B$ to a constant of mass dimension $+1$, but no such object can be constructed from the spinors and scales at hand. Thus, we set $B=0$. In summary, the Einstein  mode corresponds to:
\be\label{nhEin}
\mbox{Einstein:} \qquad \{B,\,B^{\alpha}\}=\left\{0,\, \kappa\,\lambda^{\alpha}\right\}\,, \quad B_{A}=\kappa\left(\begin{array}{c}
                                                                                                               0 \\
                                                                                                               \lambda^{\alpha} 
                                                                                                               \end{array}\right)\,, 
\ee
leading to the usual negative helicity Einstein graviton: 
\be  \label{2222}
\Psi_{\alpha\beta\gamma\delta}=\kappa\,\lambda_{\alpha}\lambda_{\beta}\lambda_{\gamma}\lambda_{\delta}\,\e^{\im\,k\cdot x} \ . \ee
%
There are two other linearly independent solutions, which solve the 4-derivative equations of motion. The first of these is specified by the choice of a constant, mass dimension $-\frac{1}{2}$ spinor $a^{\alpha}$ which satisfies $\la a\,\lambda\ra\neq 0$ and carries the opposite little group weight to $\lambda_{\alpha}$. This state is referred to as a `spin-1' state:
\be\label{nhS1}
\mbox{Spin-1:} \qquad \{B,\,B^{\alpha}\}=\left\{0,\, a^{\alpha}\right\}\,, \quad B_{A}=\left(\begin{array}{c}
                                                                                              -\frac{\im}{2}\,\tilde{\lambda}_{\dot\alpha}\,\la a\,\lambda\ra \\
                                                                                              a^{\alpha}
                                                                                             \end{array}\right).
\ee
Taking into account the mass dimension and little group weight of $a^{\alpha}$, it is easy to see that the field strength
\be\label{nhsfs}
\Psi^{\prime}_{\alpha\beta\gamma\delta}=a_{(\alpha}\lambda_{\beta}\lambda_{\gamma}\lambda_{\delta)}\,\e^{\im\,k\cdot x}\,
\ee
corresponds to a helicity $-1$ mode. Note that although this is a solution of the Bach equation (without solving any lower-derivative equations), it is a suitable state for scattering since it is purely oscillatory.

The third linearly independent solution is the growing state:
\be\label{nhGh}
\mbox{Growing:} \qquad \{B,\,B^{\alpha}\}=\left\{[\tilde{\lambda}\,\tilde{\beta}],\,x^{\alpha\dot\beta}\tilde{\beta}_{\dot\beta}\right\}\,, \quad B_{A}= \left(\begin{array}{c}
                                                                                              \tilde{\beta}_{\dot\alpha}-\frac{\im}{2}\la\lambda|x|\tilde{\beta}]\,\tilde{\lambda}_{\dot\alpha}\\
                                                                                             x^{\alpha\dot\beta}\tilde{\beta}_{\dot\beta}
                                                                                            \end{array}\right)\,,
\ee
which is specified by a choice of a constant, mass dimension $+\frac{1}{2}$ spinor $\tilde{\beta}_{\dot\alpha}$. Since the linearised field strength  grows   linearly with $x$: 
\be\label{nhgfs}
\Psi^{\mathrm{g}}_{\alpha\beta\gamma\delta}=\lambda_{(\alpha}\lambda_{\beta}\lambda_{\gamma}\,x_{\delta)}{}^{\dot\alpha}\tilde{\beta}_{\dot\alpha}\,\e^{\im\,k\cdot x}\,,
\ee
such  growing states must be excluded from the set external states
in defining an  S-matrix.  

\medskip

The construction for the positive helicity sector proceeds in a similar fashion. In this case, the conformal gravity field is constructed from a helicity $+\frac{3}{2}$ Rarita-Schwinger field via a \emph{helicity raising operator} $A^{A}$ 
\be\label{ph1}
\widetilde{\Gamma}^{A}{}_{\dot\beta\dot\gamma\dot\delta}=A^{A}\,\tilde{\lambda}_{\dot\beta}\tilde{\lambda}_{\dot\gamma}\tilde{\lambda}_{\dot\delta}\,\e^{\im\,k\cdot x}\,, \qquad \qquad A^{A} =(\tilde{A}^{\dot\alpha},\,A_{\alpha}) \ . 
\ee
This field obeys a constraint
\be\label{ph2}
\widetilde{C}_{A}\,\widetilde{\Gamma}^{A}{}_{\dot\beta\dot\gamma\dot\delta}=0\,,
\ee
in terms of the momentum space differential operator
\be\label{dopt}
\widetilde{C}_{A}:=\left(\tilde{\lambda}_{\dot\alpha},\, -\im \frac{\partial}{\partial\lambda_{\alpha}}\right)\,.
\ee
With some elementary assumptions, this translates into a condition on the components of $A^{A}$
\be\label{ph3}
\im\,\tilde{\lambda}_{\dot\alpha}\,\tilde{A}^{\dot\alpha}+\frac{\partial A_{\alpha}}{\partial\lambda_{\alpha}}=0\,,
\ee
which can be solved for $A_{\alpha}$:
\be\label{ph4}
A_{\alpha}=\frac{\partial \tilde{A}}{\partial\lambda^{\alpha}}-\im\,\frac{\lambda_{\alpha}}{2}\,\tilde{\lambda}_{\dot\alpha}\,\tilde{A}^{\dot\alpha}\,.
\ee
As expected, there are three d.o.f. parametrized by $\{\tilde{A},\,\tilde{A}^{\dot\alpha}\}$. The various objects appearing in the definition of the helicity raising operator have mass dimensions
\be\label{ph5}
[\tilde{A}^{\dot\alpha}]=-\frac{1}{2}\,, \qquad [A_{\alpha}]=\frac{1}{2}\,, \qquad [\tilde{A}]=1\,,
\ee
as dictated by conformal invariance.

The decomposition of these three d.o.f. into distinct states is accomplished in the same way as in  the negative helicity case. The result is one Einstein, one spin-1  and one growing mode:
\be\label{phEin}
\mbox{Einstein:} \qquad \{\tilde{A},\,\tilde{A}^{\dot\alpha}\}=\left\{0,\,\kappa\tilde{\lambda}^{\dot\alpha}\right\}\,, \quad A^{A}=\kappa\left(\begin{array}{c}
                                                                                                                                                      \tilde{\lambda}^{\dot\alpha} \\
                                                                                                                                                      0
                                                                                                                                                      \end{array}\right)\,,
\ee
\be\label{phS1}
\mbox{Spin-1:} \qquad \{\tilde{A},\,\tilde{A}^{\dot\alpha}\}=\left\{0,\,\tilde{a}^{\dot\alpha}\right\}\,, \quad A^{A}=\left(\begin{array}{c}
                                                                                                                                       \tilde{a}^{\dot\alpha}  \\
                                                                                                                                       -\frac{\im}{2}\lambda_{\alpha}\,[\tilde{a}\,\tilde{\lambda}]
                                                                                                                                                      \end{array}\right)\,.
\ee
\be\label{phGh}
\mbox{Growing:} \qquad \{\tilde{A},\,\tilde{A}^{\dot\alpha}\}=\left\{\la\lambda\,\beta\ra,\,x^{\gamma\dot\alpha}\beta_{\gamma}\right\}\,, \quad A^{A}=\left(\begin{array}{c}
                                                                                                                                       x^{\gamma\dot\alpha}\beta_{\gamma}  \\
                                                                                                                                       \beta_{\alpha}-\frac{\im}{2}\,\la\beta|x|\tilde{\lambda}]\,\lambda_{\alpha}
                                                                                                                                                      \end{array}\right)\,.
\ee
The constant spinors appearing in this decomposition have mass dimensions
\be\label{ph6}
[\beta_{\alpha}]=\frac{1}{2}\,, \qquad [\tilde{a}^{\dot\alpha}]=-\frac{1}{2}\,,
\ee
and the corresponding field strengths are:
\be\label{phfs}
\widetilde{\Psi}_{\dot\alpha\dot\beta\dot\gamma\dot\delta}=\kappa\,\tilde{\lambda}_{\dot\alpha}\tilde{\lambda}_{\dot\beta}\tilde{\lambda}_{\dot\gamma}\tilde{\lambda}_{\dot\delta}\,\e^{\im\,k\cdot x}\,, \qquad
\widetilde{\Psi}^{\prime}_{\dot\alpha\dot\beta\dot\gamma\dot\delta}=\tilde{a}_{(\dot\alpha}\tilde{\lambda}_{\dot\beta}\tilde{\lambda}_{\dot\gamma}\tilde{\lambda}_{\dot\delta)}\,\e^{\im\,k\cdot x}\,,
\ee
\begin{equation*}
 \widetilde{\Psi}^{\mathrm{g}}_{\dot\alpha\dot\beta\dot\gamma\dot\delta}=\tilde{\lambda}_{(\dot\alpha}\tilde{\lambda}_{\dot\beta}\tilde{\lambda}_{\dot\gamma}\,x^{\alpha}{}_{\dot\delta)}\beta_{\alpha}\,\e^{\im\,k\cdot x}\,.
\end{equation*}
It is easy to see that in each case, the positive helicity fields are simply the helicity conjugates of their negative helicity counterparts.

One could worry that the constant spinors $a_{\alpha},\tilde{a}_{\dot\alpha}, \beta_{\alpha}, \tilde{\beta}_{\dot\alpha}$ correspond to additional d.o.f., but this is not the case. Indeed, the conditions
\begin{equation*}
 \la a\,\lambda\ra\neq0\,, \quad \la\beta\,\lambda\ra\neq0\,, \quad [\tilde{a}\,\tilde{\lambda}]\neq 0\,, \quad [\tilde{\beta}\,\tilde{\lambda}]\neq 0\,,
\end{equation*}
already fix each spinor up to a scale. The $\beta_{\alpha},\tilde{\beta}_{\dot\alpha}$ spinors can then be fixed as
\begin{equation}
 \beta_{\alpha}=a_{\alpha}\,\la\beta\,\lambda\ra\,, \qquad \tilde{\beta}_{\dot\alpha}=\tilde{a}_{\dot\alpha}\,[\tilde{\beta}\,\tilde{\lambda}]\,,
\end{equation}
with scalar products  $\la\beta\,\lambda\ra$, $[\tilde{\beta}\,\tilde{\lambda}]$ setting the overall scale.

\medskip

In summary, the twistor-spinor formalism provides an easy way to capture all of the d.o.f. of conformal gravity in a way that is manifestly conformally invariant. The helicity raising/lowering operators serve as `polarizations': by selecting which d.o.f. appear in the operator, we can single out the individual states of the on-shell theory. 


\subsection{Polarizations and double copy}

To enable comparison with the results obtained in the twistor-spinor formalism, it will be useful to have expressions for the scattering states of conformal gravity in terms of the standard metric perturbation  $h_{ab}$. Since the  growing 
states are excluded from this class, this entails finding polarization tensors for the Einstein  graviton and spin-1 modes of the conformal gravity field. To do this, we write
\begin{equation}
 h_{ab}=\varepsilon_{ab}\,\e^{\im\,k\cdot x}\,,
\end{equation}
for some constant polarization $\varepsilon_{ab}$, and assume that the mass dimension $-\frac{1}{2}$ spinors $a^{\alpha}$, $\tilde{a}^{\dot\alpha}$ appearing in \eqref{nhS1} and \eqref{phS1} are normalized to obey
\be\label{spinnorm}
\la a\,\lambda\ra=1=[\tilde{a}\,\tilde{\lambda}]\,.
\ee
This normalization can be viewed as  expressing 
$ a^{\alpha}=\frac{\omega^{\alpha}}{\la\omega\,\lambda\ra}\,$
in terms of a dimensionless spinor $\omega^{\alpha}$.
Then the polarization tensors for negative and positive helicity Einstein modes are given respectively by:
\be\label{pole1}
\varepsilon^{(-2)}_{\alpha\dot\alpha\beta\dot\beta}=\lambda_{\alpha}\lambda_{\beta}\,\tilde{a}_{\dot\alpha}\tilde{a}_{\dot\beta}\,, \qquad \varepsilon^{(+2)}_{\alpha\dot\alpha\beta\dot\beta}=\tilde{\lambda}_{\dot\alpha}\tilde{\lambda}_{\dot\beta}\,a_{\alpha}a_{\beta}\,,
\ee
which are the usual expressions for Einstein graviton polarizations in the spinor helicity formalism. It is a straightforward exercise to calculate the linearised curvature tensors associated with these polarizations,
\be\label{lincurv1}
R_{abcd}^{(-2)}=\epsilon_{\dot\alpha\dot\beta}\,\epsilon_{\dot\gamma\dot\delta}\,\lambda_{\alpha}\lambda_{\beta}\lambda_{\gamma}\lambda_{\delta}\,\e^{\im\,k\cdot x}\,, \qquad R^{(+2)}_{abcd}=\epsilon_{\alpha\beta}\,\epsilon_{\gamma\delta}\,\tilde{\lambda}_{\dot\alpha}\tilde{\lambda}_{\dot\beta}\tilde{\lambda}_{\dot\gamma}\tilde{\lambda}_{\dot\delta}\,\e^{\im\,k\cdot x}\,,
\ee
confirming that they correspond to negative and positive helicity Einstein gravitons. Note that all dependence on the constant spinors $a^{\alpha}$, $\tilde{a}^{\dot\alpha}$ drops out at the level of the gauge invariant field strengths, as required in the Einstein sector.

As for the spin-1 sector, we can take the same polarization tensors as those used in~\cite{Berkovits:2004jj}:
\be\label{pols1}
\varepsilon^{(-1)}_{\alpha\dot\alpha\beta\dot\beta}=\lambda_{(\alpha} a_{\beta)}\,\tilde{a}_{\dot\alpha}\tilde{a}_{\dot\beta}\,, \qquad \varepsilon^{(+1)}_{\alpha\dot\alpha\beta\dot\beta}=\tilde{\lambda}_{(\dot\alpha} \tilde{a}_{\dot\beta)}\,a_{\alpha}a_{\beta}\,.
\ee
For the negative helicity mode, this leads to a linearised curvature
\be\label{lincurv2}
R_{abcd}^{(-1)}=\left(\epsilon_{\dot\alpha\dot\beta}\epsilon_{\dot\gamma\dot\delta}\,a_{(\alpha}\lambda_{\beta}\lambda_{\gamma}\lambda_{\delta)} +\tfrac{1}{2}\epsilon_{\dot\alpha\dot\beta}\epsilon_{\gamma\delta}\lambda_{\alpha}\lambda_{\beta}\,\tilde{a}_{(\dot\gamma}\tilde{\lambda}_{\dot\delta)}+\tfrac{1}{2}\epsilon_{\alpha\beta}\epsilon_{\dot\gamma\dot\delta}\lambda_{\gamma}\lambda_{\delta}\,\tilde{a}_{(\dot\alpha}\tilde{\lambda}_{\dot\beta)}\right)\e^{\im\,k\cdot x}\ . 
\ee
The first term is a linearised ASD Weyl spinor, corresponding to the desired behaviour of \eqref{nhsfs}, while the other two terms are contributions from a linearised Ricci tensor
\be\label{linRic}
\Phi_{\alpha\beta}^{\dot\alpha\dot\beta}=\frac{1}{2}\,\lambda_{\alpha}\lambda_{\beta}\,\tilde{a}^{(\dot\alpha}\tilde{\lambda}^{\dot\beta)}\,\e^{\im k\cdot x}\,.
\ee
These Ricci tensor contributions can be removed by a conformal transformation, and will thus decouple from any scattering amplitude calculations.

\medskip

In~\cite{Johansson:2017srf}, it was shown that certain `non-minimal' conformal gravities obey \emph{double copy}, in the sense that kinematic numerators of the theory's scattering amplitudes can be constructed by tensoring together kinematic numerators from two different gauge theories~\cite{Bern:2008qj,Bern:2010ue}. The two gauge theories which form the basis for the conformal supergravity double copy are Yang-Mills theory and a gauge theory with four-derivative equations of motion and a coupling constant of mass dimension $+1$. The kinetic term of this latter theory is schematically $(DF)^2$, where $D$ is the gauge covariant derivative and $F$ is the field strength. This $(DF)^2$ gauge theory has an ambitwistor string description~\cite{Azevedo:2017lkz} and also plays an interesting role in the construction of scattering amplitudes in heterotic and bosonic string theory~\cite{Azevedo:2018dgo}. 

There are important differences between non-minimal conformal gravity and the standard `minimal' conformal gravity we study\footnote{In particular, non-minimal conformal gravities have additional scalars which couple to the graviton through an arbitrary function, cf.~\cite{Fradkin:1983tg,Fradkin:1985am,Berkovits:2004jj,Butter:2016mtk,Tseytlin:2017qfd}.}, but at the linearised level there is no distinction and we expect some remnant of the double copy to be visible in the polarization data. Indeed, in double copy the polarizations of the gravitational theory should be expressible as symmetric products of the polarizations in the appropriate gauge theories. In Yang-Mills theory, one has only the negative and positive helicity gluon polarization vectors, which are given in spinor helicity form by:
\be\label{glupol}
e^{(-1)}_{\alpha\dot\alpha}=\lambda_{\alpha}\,\tilde{a}_{\dot\alpha}\,, \qquad e^{(+1)}_{\alpha\dot\alpha}=\tilde{\lambda}_{\dot\alpha}\,a_{\alpha}\,.
\ee
It is easy to see that taking symmetric squares of these polarization vectors generates the positive and negative helicity Einstein graviton polarizations \eqref{pole1} (as well as two scalar polarizations, corresponding to a dilaton and axion, as expected from the double copy). But what about the polarization data corresponding to the four-derivative gauge theory of~\cite{Johansson:2017srf}?
The linearised equations of motion for this four-derivative theory are
\be\label{4g1}
\Box \partial_{a}\,F^{ab}=0\,,
\ee
where $F^{ab}$ is the field strength of a gauge potential. 
Clearly, gluons are a consistent subsector of solutions to these equations, so we get another copy of the gluon polarizations \eqref{glupol}. There are also two spin-1 growing modes (one each of negative and positive helicity): $A^{\mathrm{g}}_{a}\sim \B_{a} n\cdot x\,\e^{\im\,k\cdot x}$, for some polarization $\B_{a}$ and time-like vector $n^{a}$; such modes are excluded from the set of acceptable scattering states.

Finally, the theory also includes purely oscillatory solutions with a polarization 
\be\label{scalpol}
e^{(0)}_{\alpha\dot\alpha}=a_{\alpha}\,\tilde{a}_{\dot\alpha}\,.
\ee
The little group weight of this polarization indicates that it corresponds to a scalar degree of freedom; this is confirmed by computing the linear field strength associated with \eqref{scalpol},
\be\label{sFs}
F_{ab}^{(0)}=\im\left(\epsilon_{\dot\alpha\dot\beta}\,\lambda_{(\alpha}\,a_{\beta)} + \epsilon_{\alpha\beta}\,\tilde{\lambda}_{(\dot\alpha}\,\tilde{a}_{\dot\beta)}\right)\e^{\im\,k\cdot x}\,,
\ee
which is helicity zero, having both SD and ASD parts (of equal magnitude).

From this, we conclude that there are five on-shell d.o.f. in the $(DF)^2$ theory. The spectrum consists of: positive and negative helicity (Yang-Mills) gluons; positive and negative helicity spin-1 growing modes; and a scalar which obeys the four-derivative equations of motion in a strict sense. Taking symmetric products between the Yang-Mills polarizations \eqref{glupol} and the polarizations of the scattering states in $(DF)^2$ theory gives precisely the polarizations \eqref{pole1}, \eqref{pols1} for the scattering states in conformal gravity (plus some expected scalars), as dictated by the double copy.


\section{3-point amplitudes in  conformal gravity}
\label{CGAmps}

The twistor-spinor representation of scattering states for conformal gravity enables us to give a compact expression for all of the tree-level 3-point amplitudes in that theory. In two-derivative theories, Poincar\'e covariance means that a 3-point amplitude is uniquely fixed by specifying the helicities of the external particles~\cite{Benincasa:2007xk}. The same is \emph{not} true of a higher-derivative theory such as conformal gravity. The notion of an external scattering state of positive/negative helicity is not unique: one must additionally specify whether the state being scattered is an Einstein graviton or a spin-1 field. In terms of the twistor-spinor formalism, this is captured by the choice of helicity lowering/raising operator, $B_{A}$ in \rff{nh1} or $A^{A}$ in \rff{ph1}. 

Therefore, we expect any formula for 3-point amplitudes to depend explicitly on these helicity raising/lowering operators for each particle. Indeed, one can think of these operators as higher-derivative `polarization' data for the external states in a scattering process. Amplitudes for the specific helicity states of conformal gravity are obtained by making explicit choices for these polarizations.

The only potentially non-vanishing 3-point amplitudes involve two positive helicity and one negative helicity external fields ($\overline{\mbox{MHV}}$) or two negative and one positive helicity external fields (MHV). This follows from the integrability of the self-duality equations. In the $\overline{\mbox{MHV}}$ configuration, momentum conservation dictates that all un-dotted momentum spinors are proportional ($\lambda_1\propto\lambda_2\propto\lambda_3$), so the amplitude is a function only of the dotted momentum spinors~\cite{Witten:2003nn,Benincasa:2007xk}. The opposite is true of the MHV configuration, and the two should be related by complex conjugation (i.e., exchanging positive helicity for negative helicity).

\medskip

The $\overline{\mbox{MHV}}$ 3-point amplitude of conformal gravity is given by:
\be\label{cgmb1} \cM_{3}=
\varepsilon\Big[A_{2}\cdot \widetilde{C}_{3}\Big(B_{1}\cdot A_{3}\,\frac{[2\,3]^{4}}{[1\,2]\,[3\,1]^2}\Big) + A_{3}\cdot \widetilde{C}_{2}\Big(B_{1}\cdot A_{2}\,\frac{[2\,3]^{4}}{[1\,2]^2\,[3\,1]}\Big)\Big]\; \delta^{(4)}\!\Big(\sum_{i=1}^{3}\lambda_{i}\tilde{\lambda}_{i}\Big)\,,
\ee
where $\varepsilon$ is the dimensionless coupling constant, and particle 1 has been chosen to have negative helicity whilst 2 and 3 have positive helicity. The helicity raising/lowering operators for each particle can take any of the allowed scattering state forms; namely \eqref{phEin} or \eqref{phS1} for the raising operators, and \eqref{nhEin} or \eqref{nhS1} for the lowering operators. The differential operator $\widetilde{C}_{i\,A}$ is given by \eqref{ph2}, and acts on everything to its right, including the momentum conserving delta function. 

This formula can be derived directly from a formulation of conformal gravity in twistor space~\cite{Mason:2005zm,Adamo:2013tja,Adamo:2013cra}, which is summarized in appendix~\ref{PTder}. Alternatively, we can simply posit \eqref{cgmb1} and then check that it is correct. It is easy to see that \eqref{cgmb1} passes some basic consistency tests. The formula is linear in each of the external helicity raising/lowering operators, as expected. Further, combinations $A_{i}\cdot\widetilde{C}_{j}$ and $B_{i}\cdot A_{j}$ have mass dimension zero, so the coefficient of the coupling constant and momentum conserving delta function 
 has mass dimension $+1$, as required for a conformally invariant theory.

The first substantive check on \eqref{cgmb1} comes by evaluating it when the three external particles are all Einstein gravitons. Plugging in \eqref{phEin} and \eqref{nhEin} results in:
\be\label{cgmb2}
\cM_{3}(1^{-},2^{+},3^{+})=0\,.
\ee
This is the expected result: the embedding of Einstein gravity into conformal gravity ensures that the tree-level S-matrix of Einstein states in conformal gravity vanishes on a flat background~\cite{Maldacena:2011mk,Adamo:2013cra}.  

\medskip

But what about more general scattering configurations? In particular, we are free to scatter any combination of Einstein gravitons and spin-1 modes. It is straightforward to show that the only non-vanishing results are in the configurations:
\be\label{cgmb5}
\cM_{3}(a_{1},2^{+},3^{+})=\varepsilon\,\kappa^{2}\,\frac{[2\,3]^{5}\,\la a_{1}\,1\ra}{[1\,2]\,[3\,1]}\,
\delta^{(4)}\!\Big(\sum_{i=1}^{3}\lambda_{i}\tilde{\lambda}_{i}\Big)\,,
\ee
\be\label{cgmb6}
\cM_{3}(a_{1},\tilde{a}_{2},3^{+})=\varepsilon\,\kappa\,\frac{[2\,3]^{4}\,\la a_{1}\,1\ra\,[\tilde{a}_{2}\,2]}{[1\,2]^{2}}\,\delta^{(4)}\!\Big(\sum_{i=1}^{3}\lambda_{i}\tilde{\lambda}_{i}\Big)\,,
\ee
with all others vanishing as a result of momentum conservation. Here, 
$a_i$ and $\tilde a_i$   are constant spinors parametrizing spin-1 modes in \rff{nhS1} and \rff{phS1}. The first of these \eqref{cgmb5} is anti-symmetric under the interchange of the two positive helicity Einstein gravitons: $\cM_{3}(a_1,2^{+},3^{+})=-\cM_{3}(a_1,3^{+},2^{+})$, which means that the amplitude is, in fact,  zero by crossing-symmetry.

Therefore, it is only the amplitude  \eqref{cgmb6}  that  is actually non-vanishing. Note that 
$\la a_{1}\,1\ra=1$ and $[\tilde{a}_{2}\,2]=1$ thanks to the initial normalization of the external wave-functions, meaning that explicit dependence on the constant spinors appearing in the helicity raising/lowering operators drops out of the gauge-invariant scattering amplitude:
\be\label{cgmb6*}
\cM_{3}(a_{1},\tilde{a}_{2},3^{+})=\varepsilon\,\kappa\,\frac{[2\,3]^{4}}{[1\,2]^{2}}\,\delta^{(4)}\!\Big(\sum_{i=1}^{3}\lambda_{i}\tilde{\lambda}_{i}\Big)\,.
\ee
Not surprisingly, the kinematic part of this expression matches the expected result for $\overline{\mbox{MHV}}$ amplitudes of helicity $(-1,+1,+2)$ fields~\cite{Benincasa:2007xk}. This is simply the vector-vector-graviton amplitude of Einstein-Maxwell theory, as expected from the two-derivative formulation conformal gravity given by \rff{2}.

This result (as well as the vanishing of all other external configurations for the $\overline{\mbox{MHV}}$ amplitude) has been confirmed by direct calculation from the conformal gravity action on space-time, using on-shell polarizations \eqref{pole1}, \eqref{pols1}. This proves that \eqref{cgmb1} is correct.

The helicity-conjugate MHV amplitudes are captured in the obvious way by  the following analogue of \rff{cgmb1} 
\be\label{cgmhv1}
\cM_{3} =\varepsilon\Big[B_{2}\cdot C_{3}\Big(A_{1}\cdot B_{3}\,\frac{\la2\,3\ra^{4}}{\la1\,2\ra\,\la3\,1\ra^2}\Big) + B_{3}\cdot C_{2}\Big(A_{1}\cdot B_{3}\,\frac{\la2\,3\ra^{4}}{\la1\,2\ra^2\,\la3\,1\ra}\Big)\Big]\; \delta^{(4)}\!\Big(\sum_{i=1}^{3}\lambda_{i}\tilde{\lambda}_{i}\Big)\,,
\ee
with $C_{i}^{A}$ the differential operator \eqref{dop} acting on the on-shell momenta of particle $i$. When evaluated on specific configurations for the external states, the only non-vanishing result is the helicity conjugate of \eqref{cgmb6*}.

\medskip

Observe that \eqref{cgmb1}, \eqref{cgmhv1} can also be evaluated -- formally, at least -- on growing modes. In this case, the resulting 3-point amplitudes are generically undefined in the expected sense: they are not supported on overall 4-momentum conservation due to the polynomial growth of the mode curvature, which manifests itself as derivatives of the overall momentum conserving delta function. Such highly distributional expressions for 3-point `amplitudes' in conformal gravity have previously been computed in the context of twistor-string theory~\cite{Dolan:2008gc}. However, at 3-points there exists a special degenerate configuration involving a single growing mode which results in finite, well-defined amplitudes~\cite{Brodel:2009ep}.

For the $\overline{\mbox{MHV}}$ sector, this configuration is given by a negative helicity growing mode and two positive helicity Einstein gravitons. To evaluate \eqref{cgmb1} on this configuration, we re-write the growing mode's twistor polarization as a momentum space differential operator:
\be\label{momgrow1}
B_{A}^{\mathrm{g}}\rightarrow\left(\begin{array}{c}
                                    \tilde{\beta}_{\dot\alpha}+\frac{\tilde{\beta}_{\dot\beta}}{2}\,\tilde{\lambda}_{\dot\alpha}\,\frac{\partial}{\partial\tilde{\lambda}_{\dot\beta}} \\
                                                                                             -\im\,\tilde{\beta}_{\dot\beta}\frac{\partial}{\partial k_{\alpha\dot\beta}}
                                   \end{array}\right)\,.
\ee
The un-dotted entries of this polarization, proportional to a derivative with respect to the full external momentum, are what generically lead to the breaking of 4-momentum conservation in amplitudes involving the growing mode. But in this special configuration, these contributions decouple leaving:
\be\label{strange1}
\cM_{3}(\tilde{\beta}_1, 2^{+},3^{+})=\varepsilon\,\kappa^{2}\,[1\,\tilde{\beta}_1]\,\frac{[2\,3]^{6}}{[1\,2]^{2}\,[3\,1]^{2}}\;\delta^{(4)}\!\Big(\sum_{i=1}^{3}\lambda_{i}\tilde{\lambda}_{i}\Big)\,,
\ee
which matches the `strange' 3-point amplitude of conformal gravity observed in~\cite{Brodel:2009ep}.

Note that the existence of this well-defined amplitude involving a single growing mode  
should  \emph{not}    be considered  as  evidence that growing modes should be included in the set of scattering states: in general, growing modes  lead to un-defined scattering amplitudes. Rather, it demonstrates that certain degenerate configurations may exist in which scattering a fixed number of growing modes may `accidentally' lead to finite amplitudes.\footnote{Note also that the  quantum-mechanical definition of a growing mode as an asymptotic state will still be problematic even if the multi-linear piece of the classical action evaluated on   the corresponding classical   solution 
returns some finite result.}



\section{Free fields   
and   3-point amplitudes in  CHS theory}
\label{CHSAmps}

The twistor-spinor formalism also encodes free CHS fields of any integer spin. Building on the example of conformal gravity, this allows us to write down a momentum eigenstate basis for the scattering states of CHS theory, which in turn translates into compact expressions for all 3-point tree amplitudes of the theory. To do this we work directly with the curvatures of the CHS gauge fields.
The dynamical part of a spin-$s$ CHS field is encoded in its linearised spin-$s$ Weyl tensor, $C_{a(s)b(s)}$, which is of order $s$ in derivatives of the potential, traceless on each symmetric $s$-tuple of indices, and anti-symmetric between the $s$-tuples \cite{Fradkin:1989md} (see also \ci{Marnelius:2008er,Vasiliev:2009ck}). 
The symmetries of this Weyl tensor mean that it can be decomposed into anti-self-dual and self-dual (or negative and positive helicity) parts:
\be\label{lweyldecomp}
C_{a(s)b(s)}=\epsilon_{\dot\alpha_{1}\dot\beta_{1}}\cdots\epsilon_{\dot\alpha_{s}\dot\beta_{s}}\,\Psi_{\alpha(s)\beta(s)}+\epsilon_{\alpha_1 \beta_1}\cdots\epsilon_{\alpha_s \beta_s}\,\widetilde{\Psi}_{\dot\alpha(s)\dot\beta(s)}\,,
\ee
where the spinors $\Psi_{\alpha(s)\beta(s)}$, $\widetilde{\Psi}_{\dot\alpha(s)\dot\beta(s)}$ are totally symmetric. 
In terms of these  ASD and SD parts  of the (linearised) higher spin curvature, the free field equations become
\be\label{lchs*}
\partial^{\alpha(s)\dot\alpha(s)}\,\Psi_{\alpha(s)\beta(s)}=0\,, \qquad \partial^{\beta(s)\dot\beta(s)}\,\widetilde{\Psi}_{\dot\alpha(s)\dot\beta(s)}=0\,,
\ee
generalizing the $s=2$ Bach   equations in \rff{linbach3}.


\subsection{Free fields and momentum eigenstates}

Just as the 4-derivative equations of motion for conformal gravity can be obtained from twistor-spinors with a single twistor index, the $2s$-derivative equations of motion for a free CHS field of spin $s$ can be obtained from twistor-spinors with $s-1$ twistor indices. Since this construction builds naturally on that for conformal gravity, our exposition will be much briefer than in the previous section; further details can also be found in~\cite{Haehnel:2016mlb,Adamo:2016ple}.

Let $\Gamma_{A(s-1)\,\beta(s+1)}$ be a twistor-spinor field with $s-1$ totally symmetric covariant twistor indices, and $s+1$ totally symmetric negative chirality spinor indices, which obeys the free field equation
\be\label{fchs1}
\D^{\beta\dot\beta}\,\Gamma_{A(s-1)\,\beta(s+1)}=0\,,
\ee
where $\D^{\beta\dot\beta}$ is the twistor connection \rff{1}  defined on a Minkowski background. The action of the twistor connection on multiple symmetric twistor indices follows from \eqref{LTcon2} by the Leibniz rule. Assuming that the component of $\Gamma_{A(s-1)\,\beta(s+1)}$ with all negative chirality spinor indices obeys $\Gamma^{\alpha_{1}}{}_{\alpha(s-2)\,\alpha_{1}\beta(s)}=0$,  this component can be identified with a spin-$s$ linearised ASD Weyl spinor:
\be\label{fchs2}
\Gamma_{\alpha(s-1)\,\beta(s+1)}\equiv\Psi_{\alpha(s-1)\beta(s+1)}\ ,  
\ee
with $\Psi_{\alpha(s-1)\beta(s+1)}$ a totally symmetric negative chirality spinor.

Unpacking the equation of motion \eqref{fchs1} in terms of the local twistor connection gives  a system of $s$ coupled equations for the components of $\Gamma_{A(s-1)\,\beta(s+1)}$ (cf. \rff{lcg1}--\rff{lcg4})
\begin{align}
 \partial^{\beta\dot\beta}\Psi^{\alpha(s-1)}{}_{\beta\gamma(s)}-\Gamma^{\dot\beta\alpha(s-1)}{}_{\gamma(s)} & =  0\,, \nonumber \\
 \partial^{\beta\dot\beta}\Gamma^{\alpha(s-2)}{}_{\dot\alpha\beta\gamma(s)}-\Gamma^{\dot\beta\alpha(s-2)}{}_{\dot\alpha\gamma(s)} & =  0\,, \nonumber \\
  & \vdots & \label{fchs3} \\
 \partial^{\beta\dot\beta}\Gamma^{\alpha}{}_{\dot\alpha(s-2)\beta\gamma(s)}-\Gamma^{\dot\beta\alpha}{}_{\dot\alpha(s-2)\gamma(s)} & = 0\,, \nonumber \\
 \partial^{\beta\dot\beta}\Gamma_{\dot\alpha(s-1)\beta\gamma(s)} & = 0 \,. \nonumber
\end{align}
Starting from the upper-most equation, each of these relations can be fed into the one below it, until all components of $\Gamma_{A(s-1)\beta(s+1)}$ except for $\Psi_{\alpha(s-1)\beta(s+1)}$ have been eliminated, leaving the single free field equation:
\be\label{fchs4}
\partial^{\beta(s)\dot\beta(s)}\Psi_{\alpha(s)\beta(s)}=0\,.
\ee
This is precisely the negative helicity free field equation of CHS theory in \rff{lchs*}, as desired.

For the positive helicity sector, the conjugate construction holds. That is, we begin with a twistor-spinor field $\widetilde{\Gamma}^{A(s-1)}{}_{\dot\beta(s+1)}$ obeying the obvious symmetry properties, and impose an equation of motion
\be\label{fchs5}
\D^{\beta\dot\beta}\,\widetilde{\Gamma}^{A(s-1)}{}_{\dot\beta(s+1)}=0\,.
\ee
This equation is equivalent to a system of $s$ coupled equations for the components of $\widetilde{\Gamma}^{A(s-1)}{}_{\dot\beta(s+1)}$, which imply
\be\label{fchs6}
\partial^{\beta(s)\dot\beta(s)}\widetilde{\Psi}_{\dot\alpha(s)\dot\beta(s)}=0\,,
\ee
for the totally positive chirality spinor part of $\widetilde{\Gamma}^{A(s-1)}{}_{\dot\beta(s+1)}$.

\medskip

The twistor-spinor representation of free CHS fields can now be used to provide explicit momentum eigenstate expressions for scattering states. As discussed in section~\ref{Smatrix}, a spin-$s$ CHS field has $s(s+1)$ on-shell d.o.f., which are split evenly between negative and positive helicities. These d.o.f. can then be decomposed into growing and purely oscillatory modes; at spin $s$ there are $s(s-1)$ growing and $2s$ oscillatory modes. This decomposition is discussed in detail in Appendix~\ref{DOFcount}.

Following the lesson of conformal gravity, we look for momentum eigenstate solutions which are obtained in a helicity raised/lowered manner from zero-rest-mass fields of lower spin. For a negative helicity spin-$s$ CHS field, this means finding a twistor-spinor representative of the form:
\be\label{meig1}
\Gamma_{A(s-1)\,\beta(s+1)}=B_{A(s-1)}\,\lambda_{\beta(s+1)}\,\e^{\im\,k\cdot x}\,,
\ee
where $B_{A(s-1)}$ is a generalisation of the helicity lowering operator \eqref{nh1} of conformal gravity, now an object with $s-1$ totally symmetric twistor indices. The field \eqref{meig1} obeys a constraint, derived from the twistor geometry:
\be\label{meig2}
C^{A_1}\,\Gamma_{A_1 A(s-2)\,\beta(s+1)}=0\,,
\ee
where $C^{A}$ is the momentum space operator \eqref{dop}. This translates -- with some mild assumptions akin to \eqref{nh4} -- into a set of constraints on the helicity lowering operator:
\be\label{meig3}
\frac{\partial}{\partial\tilde{\lambda}_{\dot\alpha_1}}B_{\dot\alpha_1\,A(s-2)}+\im\,\lambda_{\alpha_1}\,B^{\alpha_1}{}_{A(s-2)}=0\,,
\ee
which must hold for all values of the remaining $s-2$ twistor indices. 

These relations allow us to determine all of the components of $B_{A(s-1)}$ in terms of a set of negative chirality symmetric spinors:
\be\label{hlow1}
B_{A(s-1)} \quad \leftrightarrow \quad \left\{B,\,B^{\alpha},\,B^{\alpha_1\alpha_2},\,\ldots,\,B^{\alpha(s-1)}\right\}\,.
\ee
Sure enough, there are $\frac{s(s+1)}{2}$ d.o.f. in this set of symmetric spinors, matching the count for the negative helicity sector of CHS theory. The highest-rank spinor, $B^{\alpha(s-1)}$ encodes the CHS curvature via
\be\label{meig4}
\Psi_{\alpha(s)\beta(s)}=B_{(\alpha(s-1)}\,\lambda_{\alpha_s)}\,\lambda_{\beta(s)}\,\e^{\im\,k\cdot x}\,,
\ee
with $B^{\alpha(s-1)}$ having mass dimension $\frac{1-s}{2}$. The remaining components of $B_{A(s-1)}$ are fixed by the set $\{B,\ldots,B^{\alpha(s-1)}\}$ through the relations
\be\label{hlow2}
B_{\dot\alpha(k)}{}^{\alpha(s-k-1)}=\sum_{|I|=0}^{k}\left(-\frac{\im}{2}\right)^{|I|}\,\lambda_{\beta_I}\,\tilde{\lambda}_{(\dot\alpha_I}\,\frac{\partial^{k-|I|}B^{\beta_I \alpha(s-k-1)}}{\partial\tilde{\lambda}^{\dot\alpha_{k-I})}}\,,
\ee
for $k=0,\ldots,s-1$. The resulting components can be shown to satisfy \eqref{meig3}.

The negative helicity spin-$s$ scattering states are then obtained by making choices of $B^{\alpha(s-1)}$ which are constant (in position space), leading to purely oscillatory fields. Of course, these choices are not arbitrary: the resulting spinor field must satisfy the negative helicity field equation \eqref{fchs4}. One can show that there are $s$ linearly independent choices which are purely oscillatory and solve the field equation (this also follows from the degree of freedom counting in appendix~\ref{DOFcount}). Labelling this family of solutions by an integer $h=1,\ldots,s$, the helicity raising operator components and spinor fields are:
\be\label{meig5}
B^{\alpha(s-1)}_{h}=\kappa_{h}\,a^{(\alpha(s-h)}\,\lambda^{\alpha(h-1))}\,, \qquad \Psi^{(-h)}_{\alpha(s)\beta(s)}=\kappa_{h}\,a_{(\alpha(s-h)}\,\lambda_{\alpha(h)}\,\lambda_{\beta(s)\,)}\,\e^{\im\,k\cdot x}\,,
\ee
where $\kappa_{h}$ is a coupling constant of dimension $1-h$ (by definition,  $\kappa_1=1$), and $a^{\alpha(s-h)}$ is a totally symmetric constant spinor of dimension $\frac{h-s}{2}$ which satisfies $a_{\alpha(s-h)}\lambda^{\alpha(s-h)}\neq 0$. With the proviso that $a_{\alpha(s-h)}$ scales with the opposite little group weight to $\lambda_{\alpha(s-h)}$, it follows that $\Psi^{(-h)}_{\alpha(s)\beta(s)}$ has helicity $-h$.

The helicity lowering components associated with growing modes are also easily deduced from appendix~\ref{DOFcount}, but since these are not admissible scattering states we will not 
 treat them explicitly here. The positive helicity states are derived in a similar manner, with a helicity raising operator $A^{B(s-1)}$ determining a spin-$s$ CHS field via:
\be\label{meig6}
\widetilde{\Gamma}^{B}{}_{\dot\alpha(s+1)}=A^{B(s-1)}\,\tilde{\lambda}_{\dot\alpha(s+1)}\,\e^{\im\,k\cdot x}\,, \qquad \D^{\alpha\dot\alpha}\widetilde{\Gamma}^{B(s-1)}{}_{\dot\alpha(s+1)}=0\,,
\ee
subject to the constraint 
\be\label{meig7}
\im\,\tilde{\lambda}_{\dot\beta_1}\,A^{\dot\beta_1 B(s-2)}+\frac{\partial}{\partial\lambda_{\beta_1}}A_{\beta_1}{}^{B(s-2)}=0\,.
\ee
As in the negative helicity case, this allows us to solve for all of the components of $A^{B(s-1)}$ in terms of a set of symmetric spinors,
\be\label{hrai1}
A^{B(s-1)} \quad \leftrightarrow \quad \left\{\tilde{A},\,\tilde{A}^{\dot\beta},\,\tilde{A}^{\dot\beta_1 \dot\beta_2},\,\ldots,\,\tilde{A}^{\dot\beta(s-1)}\right\}\,,
\ee
via the relations
\be\label{hrai2}
A_{\beta(k)}{}^{\dot\beta(s-k-1)}=\sum_{|I|=0}^{k} \left(-\frac{\im}{2}\right)^{|I|}\,\tilde{\lambda}_{\dot\alpha_I}\,\lambda_{(\beta_I}\,\frac{\partial^{k-|I|}\tilde{A}^{\dot\alpha_I \dot\beta(s-k-1)}}{\partial\lambda^{\beta_{k-I})}}\,.
\ee
The $s$ independent scattering states, labelled by their integer helicity $h=1,\ldots,s$ are then given by:
\be\label{meig8}
\tilde{A}^{\dot\beta(s-1)}_{h}=\kappa_{h}\,\tilde{a}^{(\dot\beta(s-h)}\,\tilde{\lambda}^{\dot\beta(h-1))}\,, \qquad \widetilde{\Psi}^{(h)}_{\dot\alpha(s)\dot\beta(s)}=\kappa_{h}\,\tilde{a}_{(\dot\alpha(s-h)}\,\tilde{\lambda}_{\dot\alpha(h)}\,\tilde{\lambda}_{\dot\beta(s)\,)}\,\e^{\im\,k\cdot x}\,,
\ee
with $\tilde{a}^{\dot\beta(s-h)}$ a totally symmetric constant spinor of dimension $\frac{h-s}{2}$ obeying $\tilde{a}_{\dot\beta(s-h)}\tilde{\lambda}^{\dot\beta(s-h)}\neq 0$.


\subsection{3-point amplitudes of CHS theory}

The twistor-spinor representation for CHS fields allows us to write expressions for \emph{all} of the tree-level 3-point amplitudes of the theory. Once again, these 3-point amplitudes are either $\overline{\mbox{MHV}}$ or MHV. There are only mild constraints on the allowed spins of the external states: if particles 2 and 3 have the same helicity sign (positive for $\overline{\mbox{MHV}}$ and negative for MHV), then the spins of the three external states must obey:
\be\label{extspins}
s_{1}\geq s_{2},s_{3}\,, \qquad s_{1}\leq s_{2}+s_{3}\,.
\ee
For spins satisfying these constraints, the 3-point $\overline{\mbox{MHV}}$ amplitude reads:
\begin{multline}\label{chs3p1}
\cM_{3}\ \sim\  \cN^{(\mathbf{s})}\left[\frac{(s_{2}-1)!}{(s_1 -s_3)!}\,A_{2}^{B_{K}(A_{J}}\,\tilde{C}_{3\,B_K}\left(B_{1\,A_I}\, A_{3}^{A_{I-J})}\,\frac{[2\,3]^{s_1 +2}}{[1\,2]^{s_{1}-s_{2}+1}\,[3\,1]^{s_2}}\right) \right. \\
\left.  +(-1)^{s_{2}+s_{3}-s_{1}} \frac{(s_{3}-1)!}{(s_1 -s_2)!}\,A_{3}^{B_{K}(A_{I-J-K}}\,\tilde{C}_{2\,B_K}\left(B_{1\,A_I}\,A_{2}^{A_{J+K})}\,\frac{[2\,3]^{s_{1}+2}}{[1\,2]^{s_3}\,[3\,1]^{s_{1}-s_{3}+1}}\right)\right]\,,
\end{multline}
where the spin $s_1$ particle has negative helicity, the spin $s_2$ and spin $s_3$ particles have positive helicity, multi-index labels obey
\be\label{multlabel}
|I|=s_1-1\,, \qquad |J|=s_1-s_3\,, \qquad |K|=s_2+s_3-s_1-1\,,
\ee
and 
\be\label{chsnorm}
\cN^{(\mathbf{s})}:=\frac{1}{(s_{2}+s_{3}-s_{1}-1)!}\,
\ee
is a spin-dependent normalisation constant. In \eqref{chs3p1}, we have stripped off an overall factor of the dimensionless CHS coupling constant  $\varepsilon$
 as well as an overall momentum conserving delta function. 

The only ingredients in this formula are the twistor `polarizations' of each external particle
\be\label{chstpol}
B_{1\,A(s_1-1)}\,, \quad A_{2}^{B(s_2 -1)}\,, \quad A_{3}^{B(s_3-1)}\,,
\ee
powers of the momentum space differential operator $\tilde{C}_{i\,A}$ defined by \eqref{dopt}, and contractions of the on-shell momentum spinors of each external particle.\foot{Recall that only `anti-holomorphic' square bracket contractions can appear in the $\overline{\mbox{MHV}}$ amplitude, as all angle bracket contractions vanish.}
As expected, \eqref{chs3p1} has the appropriate mass dimension for a scattering amplitude in a conformally invariant theory, and is linear in each particle's twistor polarization.

The amplitudes for specific helicity configurations of the external states are read off from this formula by inserting the appropriate twistor polarizations. For instance, using \eqref{meig5}, \eqref{meig8} it is easy to see that the configuration in which each external state has helicity $\pm s_i$ leads to vanishing amplitudes:
\be\label{chs3p2}
\cM_{3}(-s_{1},s_{2},s_{3})=0\,.
\ee
This is in line with the claim that the S-matrix of  standard 
 two-derivative massless   HS  states   in 
 CHS theory vanishes~\cite{Joung:2015eny,Beccaria:2016syk,Adamo:2016ple}. 

However, it is easy to see that there are other helicity configurations which lead to non-vanishing amplitudes. Consider the family of $\overline{\mbox{MHV}}$ amplitudes with the helicity configuration $(-h_{s_1},s_{2},s_{3})$, where the negative helicity can take any of the values $-h_{s_1}=-1,\ldots,-s_1$. From \eqref{chs3p1}, it can be seen that 
\be\label{chs3p3}
\cM_{3}(-h_{s_1},s_2,s_3)=0\,, \qquad \qquad \forall \ h_{s_1}=2,\ldots,s_1\,,
\ee
while a non-vanishing amplitude is obtained for $h_{s_1}=1$:
\be\label{chs3p4}
\cM_{3}(-1_{s_1},s_2,s_3)=\varepsilon\,\kappa_{s_2}\,\kappa_{s_3}\,\mathcal{K}\,\frac{[2\,3]^{s_2+s_3+1}}{[1\,2]^{s_3-s_2+1}\,[3\,1]^{s_2-s_3+1}}\,\delta^{(4)}\!\left(\sum_{i=1}^{3}\lambda_{i}\tilde{\lambda}_i\right)\,.
\ee
Here, there is an overall spin-dependent numerical factor
\be\label{chsnorm2}
\mathcal{K}:=\cN^{(\mathbf{s})}\,\Big(\frac{1}{2\,\im}\Big)^{s_1-1}\Big[ (-1)^{s_1-s_3}\,\frac{(s_2 -1)!}{(s_1-s_3)!}+(-1)^{s_2}\,\frac{(s_3-1)!}{(s_1-s_2)!}\Big]\,,
\ee
and we have used the normalisation
\be\label{chsspin}
a_{1}^{\alpha(s_1 -1)}\,\lambda_{1\,\alpha(s_1-1)}=1\,,
\ee
for the dimension $\frac{1-s_1}{2}$ constant spinor $a_{1}^{\alpha(s_1-1)}$ defining the helicity $-1$ mode of the spin $s_1$ field.


\section{Scattering in 
AdS background}
\label{AdS}

The underlying conformal invariance of CHS theory suggests that it should be possible to study `scattering' of CHS fields on a background with a cosmological constant $\Lambda\neq0$. In a de Sitter ($\Lambda>0$) background, the analogue of a scattering amplitude is not uniquely defined (cf. \cite{Witten:2001kn,Strominger:2001pn,Maldacena:2002vr}). There is a mathematically consistent S-matrix propagating data from past to future infinity, but its elements are physically unobservable. Alternatively, one can define `scattering' in terms of the in-in formalism on the observable patch of dS. By contrast, in an anti-de Sitter ($\Lambda<0$) background the notion of `scattering' is uniquely defined in terms of boundary correlation functions.

For clarity, let us focus on an AdS$_4$ background with $\Lambda<0$, where the analogue of a tree-level scattering amplitude is again defined in terms of a multi-linear piece of the classical action, evaluated on solutions to the linearised equations of motion with specified boundary behaviour. It should be noted that, at tree-level, AdS amplitudes obtained in this fashion are easily related to those obtained from the in-in formalism on dS  space  by an analytic continuation~\cite{Maldacena:2002vr}. In any theory with (classical) conformal invariance, calculating such AdS `amplitudes' at tree-level is equivalent to calculating the amplitude in a flat half-space (cf. \cite{Maldacena:2011nz}). In other words, the calculation of an AdS scattering amplitude in a conformally invariant theory is the same as in flat space, up to boundary conditions.

For a higher-derivative theory like CHS, the altered boundary conditions of AdS have important consequences. First of all, our classification of scattering states on a Minkowski background used the criteria that such states lead to amplitudes supported on momentum conservation. But AdS scattering amplitudes \emph{never} manifest full momentum conservation as there is no global space-like Killing vector when $\Lambda<0$. Instead, one expects momentum conservation only in the directions parallel to the AdS boundary; conservation in the direction transverse to the boundary is replaced by a singularity in the transverse momenta (cf.   \cite{Maldacena:2002vr,Maldacena:2011nz,Raju:2012zr,Arkani-Hamed:2015bza}). This means that our definition of a scattering state in AdS should be amended to be any solution to the free equations (linearised around AdS$_4$) which leads to finite multi-linear pieces of the action, consistent with momentum conservation \emph{up to the AdS isometries}.

The second consequence is that there is a new dimensionful parameter $\Lambda$ in play which was not available in Minkowski space. As we will see, this parameter allows us to construct new solutions to the linearised equations of motion which vanish in the flat space limit.

\medskip

As above, we first  discuss  
the  case of the conformal gravity  before generalising to the full CHS theory. Consider AdS$_4$ with the metric:\foot{This slightly non-standard looking AdS$_4$  metric is just the analytic   continuation $\Lambda \to -\Lambda$ of the standard  $S^4$  metric, up to a rescaling of coordinates to take into account that $|\Lambda|= { 3R^{-2}}$ instead of ${1\ov 4} R^{-2}$  in terms of the radius $R$ of the sphere.} 
\be\label{ads1}
\d s^2 = \frac{\d x_{\alpha\dot\alpha}\,\d x^{\alpha\dot\alpha}}{(1+\Lambda\,x^2)^2}\,,
\ee
where $\Lambda<0$ is the cosmological constant.
 In these coordinates, the AdS$_4$ boundary is the hypersurface (3-sphere) $1+\Lambda x^2=0$ 
 in the affine Minkowski space charted by $x^{\alpha\dot\alpha}$. Working in the coordinates \eqref{ads1} is advantageous as there is a manifest and smooth flat space limit given by simply taking $\Lambda\rightarrow 0$. In this limit, the hypersurface $1+\Lambda x^2=0$ approaches the conformal boundary $\scri$ of Minkowski space.

In this metric, the twistor connection for AdS$_4$ reads
\be\label{adstc}
\D_{\alpha\dot\alpha}=\nabla_{\alpha\dot\alpha}+(\CA_{\alpha\dot\alpha})^{B}{}_{C}=\nabla_{\alpha\dot\alpha}+\left(\begin{array}{cc}
                                                       0 & \delta^{\gamma}_{\alpha}\,\delta^{\dot\beta}_{\dot\alpha} \\
                                                       \Lambda\,\epsilon_{\alpha\beta}\,\epsilon_{\dot\alpha\dot\gamma} & 0
                                                      \end{array}\right)\,,
\ee
where $\nabla_{\alpha\dot\alpha}$ is the Levi-Civita connection of \eqref{ads1}. The linearised Bach equations for negative and positive helicity free fields are expressed via the action of this connection on twistor-spinors as before:
\be\label{adsbeqs}
\D^{\beta\dot\beta}\Gamma_{A\beta\gamma\delta}=\D^{\beta\dot\beta}\left(\begin{array}{c}
                                                                         \gamma_{\dot\alpha \beta\gamma\delta} \\
                                                                         \Psi^{\alpha}{}_{\beta\gamma\delta}
                                                                        \end{array}\right) =0\,, \qquad
\D^{\beta\dot\beta}\widetilde{\Gamma}^{A}{}_{\dot\beta\dot\gamma\dot\delta}=\D^{\beta\dot\beta}\left(\begin{array}{c}
                                                                                                      \widetilde{\Psi}^{\dot\alpha}{}_{\dot\beta\dot\gamma\dot\delta} \\
                                                                                                      \tilde{\gamma}_{\alpha\dot\beta\dot\gamma\dot\delta}
                                                                                                     \end{array}\right)=0\,.
\ee
In terms of the  components of the twistor spinors, these equations are equivalent to
\be\label{adsbeq2}
\nabla^{\beta\dot\beta}\gamma_{\dot\alpha\beta\gamma\delta}=0\,, \qquad \nabla^{\beta\dot\beta}\Psi^{\alpha}{}_{\beta\gamma\delta}-\gamma^{\dot\beta\alpha}{}_{\gamma\delta}=0\,,
\ee
\be\label{adsbeq3}
\nabla^{\beta\dot\beta}\widetilde{\Psi}^{\dot\alpha}{}_{\dot\beta\dot\gamma\dot\delta}-\tilde{\gamma}^{\beta\dot\alpha}{}_{\dot\gamma\dot\delta}=0\,, \qquad \nabla^{\beta\dot\beta}\tilde{\gamma}_{\alpha\dot\beta\dot\gamma\dot\delta}=0\,.
\ee
As in Minkowski space, these relations can be used to eliminate the fields $\gamma_{\dot\alpha\beta\gamma\delta}$ and $\tilde{\gamma}_{\alpha\dot\beta\dot\gamma\dot\delta}$ leaving only the usual linearised Bach equations
in terms of the Weyl spinors
\be\label{adslinbach}
\nabla^{\alpha\dot\alpha}\nabla^{\beta\dot\beta}\Psi_{\alpha\beta\gamma\delta}=0=\nabla^{\alpha\dot\alpha}\nabla^{\beta\dot\beta}\widetilde{\Psi}_{\dot\alpha\dot\beta\dot\gamma\dot\delta}\,.
\ee
Although the twistor-spinor formalism carries over to AdS$_4$, our previous basis of momentum eigenstate solutions does not. To see this, it is useful to exploit the conformal flatness of AdS$_4$
to   rewrite the equations \eqref{adsbeq2} -- \eqref{adsbeq3} in affine Minkowski coordinates (cf. \cite{Penrose:1986cb}). For the negative helicity sector one finds (the positive helicity sector is given by the obvious conjugation)
\be\label{amink1}
\partial^{\beta\dot\beta}\gamma_{\dot\alpha\beta\gamma\delta}+\frac{2\,\Lambda}{1+\Lambda\, x^2}\left(x^{\beta}{}_{\dot\alpha}\,\gamma^{\dot\beta}{}_{\beta\gamma\delta}-x^{\beta\dot\beta}\,\gamma_{\dot\alpha\beta\gamma\delta}\right)=0\,,
\ee
\be\label{amink2}
\partial^{\beta\dot\beta}\Psi^{\alpha}{}_{\beta\gamma\delta}-\frac{2\,\Lambda}{1+\Lambda\,x^2}\,x^{\beta\dot\beta}\,\Psi^{\alpha}{}_{\beta\gamma\delta}=\frac{\gamma^{\dot\beta\alpha}{}_{\gamma\delta}}{1+\Lambda\,x^2}\,,
\ee
where all spinor indices are now raised and lowered with the usual Levi-Civita symbols. As in flat space, we want to classify linearly independent solutions to these coupled equations of plane wave form, encoded by a twistor-spinor
\be\label{adspw}
\Gamma_{A\beta\gamma\delta}=B_{A}\,\lambda_{\beta}\lambda_{\gamma}\lambda_{\delta}\,\e^{\im\,k\cdot x}\,,
\ee
where $B_{A}$ is a helicity lowering operator.

Of course, the 2-derivative Einstein graviton solution remains for $\Lambda<0$, although the helicity lowering operator picks up a non-trivial dotted component  (cf. \rff{nhEin})
\be\label{adsEin}
\mbox{Einstein:} \qquad \qquad B_{A}=\kappa\,\left(\begin{array}{c}
2\Lambda\,x_{\beta\dot\alpha}\,\lambda^{\beta} \\
\lambda^{\alpha}
\end{array}\right)\,. 
\ee
Similarly, the spin-1 mode is given by a deformation of its flat space form  \rff{nhS1} 
\be\label{adsS1}
\mbox{Spin-1:} \qquad \qquad B_{A}=\left(\begin{array}{c}
                                          \frac{\im}{2} \tilde{\lambda}_{\dot\alpha}\,\la\lambda\,a\ra-\Lambda\,x^{\beta\dot\alpha}\,a_{\beta} \\
                                          a^{\alpha}
                                         \end{array}\right)\,,
\ee
where $a^{\alpha}$ is the same mass-dimension $-\frac{1}{2}$ constant spinor that appeared in \eqref{nhS1}.

However, a straightforward calculation reveals that the mode \eqref{nhGh} can \emph{not} be deformed into a solution of the equations of motion \eqref{amink1}--\eqref{amink2} when $\Lambda\neq0$. Instead, we find an additional spin 2 mode:
\be\label{adsGr}
\mbox{Spin-2:} \qquad \qquad B_{A}=\Lambda\,\kappa\,\left(\begin{array}{c}
                                                            x^{\beta}{}_{\dot\alpha}\,\lambda_{\beta} \\
                                                            \frac{\im}{2}\,x^{2}\,\lambda^{\alpha}
                                                           \end{array}\right)\,.
\ee
Although this mode has quadratic polynomial dependence on the space-time coordinates, it does not make sense to refer to it as a `growing' mode in AdS$_4$. Indeed, on the AdS boundary $x^2=-\frac{1}{\Lambda}$ the curvature associated with this mode is perfectly finite.\foot{ 
Note that the  above spin-1 \rff{adsS1} and spin-2  \eqref{adsGr} modes taken 
together represent the well-known partially-massless   graviton mode \ci{Deser:1983tm}
that has only scalar gauge invariance  and thus carries $5-1=4$ effective degrees of freedom.}

At this point, it is clear that Minkowski space is a singular background from the perspective of perturbative conformal gravity: the spin two mode \eqref{adsGr} disappears in the flat space limit, where it is replaced by the growing mode \eqref{nhGh} which exists \emph{only} when $\Lambda=0$. This is a reflection of the well-known `linearization instability' of conformal gravity~\cite{Boulware:1983td}. The truncation to scattering states in Minkowski space removes precisely these problematic modes from the external states of well-defined amplitudes. However, it may be the case that the space of scattering states is enlarged away from $\Lambda=0$ to include the spin 2 states \eqref{adsGr}.

\medskip

Having obtained a basis of linearised states on AdS$_4$, one can now consider their `scattering.' We leave a more general analysis of this problem to future work, but conjecture that the 3-point amplitude expressions \eqref{cgmb1}, \eqref{cgmhv1} remain valid in AdS$_4$, up to potential boundary contributions to the AdS amplitude. To support this conjecture, we consider the scattering of three Einstein modes; the on-shell relationship between conformal gravity with Neumann boundary conditions and the (renormalized) Einstein-Hilbert action~\cite{Anderson:2001,Maldacena:2011mk} indicates that this should produce a result which is proportional to the Minkowski space scattering amplitudes of 
Einstein gravity, where the constant of proportionality is $\Lambda$.

At first, it may seem that the twistor polarization \eqref{adsEin} for an Einstein mode is unsuitable for our 3-point amplitude formulae. These are momentum space formulae, but the twistor polarization of the Einstein mode is now a function of $x$ as well as the momenta. Fortunately, this can be rectified by remembering that the twistor polarizations -- interpreted as helicity lowering/raising operators -- should be thought of as acting on momentum eigenstates. This means that we can replace the linear $x$-dependence in \eqref{adsEin} with a momentum space derivative. The resulting positive and negative helicity Einstein polarizations are then
\be\label{adsEin2}
  A^{A}=\kappa\,\left(\begin{array}{c}
                     \tilde{\lambda}^{\dot\alpha} \\
                     -2\im\,\Lambda\,\frac{\partial}{\partial\lambda^{\alpha}}
                    \end{array}\right)\,, \qquad 
B_{A}=\kappa\,\left(\begin{array}{c}
                     -2\im\,\Lambda\,\frac{\partial}{\partial\tilde{\lambda}^{\dot\alpha}} \\
                     \lambda^{\alpha}
                    \end{array}\right)\,,
\ee                    
which are suitable for evaluation in the formulae \eqref{cgmb1}, \eqref{cgmhv1}.

In the $\overline{\mbox{MHV}}$ configuration of \eqref{cgmb1}, note that
\be\label{adstcont1}\te 
A_{2}\cdot\widetilde{C}_{3}= \kappa\left([2\,3]-2\Lambda\,\left\la\frac{\partial}{\partial\lambda_{3}}\,\frac{\partial}{\partial\lambda_{2}}\right\ra\right)\,, \qquad 
B_{1}\cdot A_{3}=-2\,\im\,\Lambda\,\kappa^{2}\left(\tilde{\lambda}^{\dot\alpha}_{3}\frac{\partial}{\partial\tilde{\lambda}^{\dot\alpha}_{1}}+\lambda_{1}^{\alpha}\frac{\partial}{\partial\lambda^{\alpha}_3}\right)\,,
\ee
and similarly for contractions with different particle labels. Feeding these into \eqref{cgmb1}, one obtains
\begin{multline}\label{adsmb1}
 \cM^{\Lambda}_{3}(1^{-},2^{+},3^{+})=-2\im\,\Lambda\,\kappa^{3}\,\varepsilon\left[\left([2\,3]-2\Lambda\,\left\la\frac{\partial}{\partial\lambda_{3}}\,\frac{\partial}{\partial\lambda_{2}}\right\ra\right)\frac{[2\,3]^{5}}{[1\,2]^{2}\,[3\,1]^{2}} \right. \\
 \left. +\left([3\,2]-2\Lambda\,\left\la\frac{\partial}{\partial\lambda_{2}}\,\frac{\partial}{\partial\lambda_{3}}\right\ra\right)\frac{[2\,3]^5}{[1\,2]^2\,[3\,1]^2}\right]\,\delta^{(4)}\!\Big(\sum_{i=1}^{3}\lambda_i\tilde{\lambda}_{i}\Big)\,,
\end{multline}
since $\lambda$-derivatives in the $B\cdot A$ contractions do not contribute. The remaining differential operators act only on the overall 4-momentum delta function, which means that we can make the replacement
\be\label{deltrep}
\left\la\frac{\partial}{\partial\lambda_{3}}\,\frac{\partial}{\partial\lambda_{2}}\right\ra=\frac{[3\,2]}{2}\,
\frac{\partial}{\partial K_{\alpha\dot\alpha}}\,\frac{\partial}{\partial K^{\alpha\dot\alpha}}\,, \qquad K^{\alpha\dot\alpha}:=(\lambda_1\,\tilde{\lambda}_1 +\lambda_2\,\tilde{\lambda}_2+\lambda_3\,\tilde{\lambda}_3)^{\alpha\dot\alpha}\,.
\ee
The wave operator in the total momenta can be denoted by $\Box_{K}=
\frac{\partial}{\partial K_{\alpha\dot\alpha}}\,\frac{\partial}{\partial K^{\alpha\dot\alpha}}$.
This reduces the expression for the $\overline{\mbox{MHV}}$ amplitude on AdS$_4$ to
\be\label{adsmb2}
\cM^{\Lambda}_{3}(1^{-},2^{+},3^{+})=-4\im\,\Lambda\,\kappa^{3}\,\varepsilon\,\frac{[2\,3]^{6}}{[1\,2]^2\,[3\,1]^2}\,(1+\Lambda\Box_{K})\,\delta^{(4)}\!\Big(\sum_{i=1}^{3}\lambda_i\tilde{\lambda}_{i}\Big)\,.
\ee
Note that the leading coefficient of $\Lambda$ is precisely the flat space $\overline{\mbox{MHV}}$ 3-point amplitude of Einstein gravity, as required by the embedding of Einstein gravity inside  the  conformal gravity. Also, 
the differential operator $(1+\Lambda\Box_K)$ explicitly breaks 4-momentum conservation, as expected for AdS amplitudes. It is easy to see that this is compatible with the metric \eqref{ads1}, confirming that the Einstein gravitons within  the 
conformal gravity are good AdS scattering states. 

This gives an additional perspective on the vanishing of the amplitudes for the Einstein sector of conformal gravity in Minkowski space, which arises by taking the $\Lambda\rightarrow 0$ limit of \eqref{adsmb2}. In other words, the tree-level S-matrix of conformal gravity evaluated on Einstein states is zero in an interesting way: it is zero \emph{times} the tree-level S-matrix of 
the Einstein  theory. 
 More generally, the normalised amplitude
\be\label{adsmb3}
\frac{\im\,\cM^{\Lambda}_{3}(1^{-},2^{+},3^{+})}{4\varepsilon\,\kappa^2\,\Lambda}=\frac{[2\,3]^{6}}{[1\,2]^2\,[3\,1]^2}\,(1+\Lambda\Box_{K})\,\delta^{(4)}\!\Big(\sum_{i=1}^{3}\lambda_i\tilde{\lambda}_{i}\Big)\,,
\ee
agrees with a formula obtained for the `bulk contribution' to the 3-point $\overline{\mbox{MHV}}$ amplitude of Einstein gravity in AdS$_4$~\cite{Adamo:2012nn,Adamo:2012xe}. This `bulk contribution' does not include boundary contributions to the full AdS amplitude, but manifests a smooth flat space limit, where such boundary contributions decouple~\cite{Adamo:2015ina}.

For completeness, the MHV 3-point AdS amplitude evaluated on Einstein states is given by the obvious helicity conjugate of \eqref{adsmb2}
\be\label{adsmhv}
\cM^{\Lambda}_{3}(1^{+},2^{-},3^{-})=-4\im\,\Lambda\,\kappa^{3}\,\varepsilon\,\frac{\la2\,3\ra^{6}}{\la1\,2\ra^2\,\la3\,1\ra^2}\,(1+\Lambda\Box_{K})\,\delta^{(4)}\!\Big(\sum_{i=1}^{3}\lambda_i\tilde{\lambda}_{i}\Big)\,,
\ee
obtained by evaluating \eqref{cgmhv1} on the polarizations \eqref{adsEin2}.

\medskip

A similar phenomenon occurs for the `scattering' of general CHS fields on an
 AdS$_4$ background. Ignoring the rest of the spectrum, the two-derivative helicity $\pm s$ states of CHS theory are represented on AdS by the twistor polarizations:
\be\label{adschs1}
A^{A(s-1)}=\kappa_{s}\,\left(\begin{array}{c}
                              \tilde{\lambda}^{\dot\alpha(s-1)} \\
                              (-2\im\,\Lambda)^{s-1}\,\frac{\partial^{s-1}}{\partial\lambda^{\alpha(s-1)}}
                             \end{array}\right)\,, \qquad
B_{A(s-1)}=\kappa_{s}\,\left(\begin{array}{c}
                              (-2\im\,\Lambda)^{s-1}\,\frac{\partial^{s-1}}{\partial\tilde{\lambda}^{\dot\alpha(s-1)}} \\
                              \lambda^{\alpha(s-1)}
                             \end{array}\right)\,,
\ee
where $\kappa_s$ is the dimension $1-s$ coupling associated with a massless spin-$s$ field. Evaluating the $\overline{\mbox{MHV}}$ 3-point formula \eqref{chs3p1} on these polarizations results in:
\begin{multline}\label{adschs2}
\cM^{\Lambda}_{3}(-s_{1}, s_{2},s_{3})=\Lambda^{s_1 -1}\,\varepsilon\,\kappa_{s_1}\kappa_{s_2}\kappa_{s_3}\,\frac{\mathfrak{n}^{(\mathbf{s})}\:\: [2\,3]^{s_1+s_2+s_3}}{[1\,2]^{s_1-s_2+s_3}\,[3\,1]^{s_1+s_2-s_3}} \\ \times\,(1+\Lambda\Box_{K})^{s_2+s_3-s_1-1}\,\delta^{(4)}\!\Big(\sum_{i=1}^{3}\lambda_i\tilde{\lambda}_{i}\Big)\,,
\end{multline}
with the spin-dependent normalisation 
\be\label{adschs3}
\mathfrak{n}^{(\mathbf{s})}:=\left(1+(-1)^{s_2+s_3-s_1}\right)\,\frac{(-2\,\im)^{s_1-1}}{(s_{2}+s_{3}-s_{1}-1)!}\,\frac{(s_1-s_2+s_3-1)!\,(s_1+s_2-s_3-1)!}{(s_1-s_2)!\,(s_1-s_3)!}\,.
\ee
This matches the result found in~\cite{Adamo:2016ple} for the AdS$_4$ 3-point $\overline{\mbox{MHV}}$ amplitude, and is proportional to the flat space 3-point amplitude for massless two-derivative higher spin fields. The overall 
constant  factor  $\Lambda^{s_1-1}$ suggests that there may be a way to isolate the tree-level S-matrix of a massless higher spin theory within the AdS amplitudes of CHS theory, analogous to the embedding of the Einstein gravity inside of 
the conformal gravity.

\acknowledgments

We thank Thales Azevedo, Henrik Johansson and Tristan McLoughlin for discussions and many helpful comments on a draft. TA thanks N. Arkani-Hamed for useful conversations at the beginning of this project. AAT  is  grateful to   R. Metsaev  for   useful discussions of related questions. This work was partially supported by the STFC grant ST/P000762/1. TA is supported by an Imperial College Junior Research Fellowship.
AAT was also supported by the Russian Science Foundation grant 14-42-00047 at Lebedev Institute.


\

\appendix

\section{Helicity  structure of  conformal graviton modes}
\label{Helicity}

Here  we give  some  details concerning the claim in section \ref{2p2} that the helicity decomposition of on-shell 
conformal graviton states is gauge-dependent at the level of the (linearised) metric.

Consider $\A_{ab}$ and $\B_{ab}$  from \rff{CGcg3} 
 in  a particular frame, where 
$k^{a} = (\omega,0,0,\omega)$  and choose the unit time-like vector  as $n^a=(1,0,0,0)$.
 Helicity is then determined by the behaviour under rotations in the plane  transverse to $x^{3}$:
\[
\mathbf{R} = \begin{pmatrix} 
1&0&0&0\\ 
0&\cos\theta &\sin\theta &0\\ 
0&- \sin\theta &\cos\theta &0\\ 
0&0&0&1
\end{pmatrix} \, .
\]
Introduce the helicity basis tensors: 
\begin{align}
T^{\pm \pm} = 
\begin{pmatrix} 
0&0&0&0\\ 
0&1& \pm i&0\\ 
0&\pm i&-1&0\\ 
0&0&0&0
\end{pmatrix} \ , 
\qquad
T^{\pm} = 
\begin{pmatrix} 
0&1&\pm i&0\\ 
1&0&0&0\\ 
\pm i&0&0&0\\ 
0&0&0&0
\end{pmatrix} \ , 
\qquad
\td T^{\pm} = 
\begin{pmatrix} 
0&0&0&0\\ 
0&0&0&1\\ 
0&0&0&\pm i\\ 
0&1&\pm i&0
\end{pmatrix} \ .  \lab{a1}
\end{align}
These satisfy:
\begin{equation}
\mathbf{R}  T^{\pm \pm} \mathbf{R}^{T} = e^{\pm 2 i \theta} T^{\pm \pm}\ , 
\qquad
\mathbf{R}  T^{\pm} \mathbf{R}^{T} = e^{\pm i \theta} T^{\pm}\ , 
\qquad
\mathbf{R}  \td T^{\pm} \mathbf{R}^{T} = e^{\pm i \theta} \td T^{\pm}\ , \lab{a2}
\qquad
\end{equation}
indicating that $T^{\pm}$ and $ \td T^{\pm}$ represent helicity $\pm1$ tensors, while $T^{\pm\pm}$ are helicity $\pm2$ tensors. 

Choosing the conformal gauge  \rff{CGcg1},\eqref{CGcg4}, specializing to the above momentum frame 
 and fixing the residual gauge \eqref{CGcg5}, one finds that 
\begin{align}
\A^{ab} & = \left( A^{++}T^{++} + A^{--}T^{--}\right)^{ab}  + \left(A^{+}T^{+} + A^{-}T^{-} \right)^{ab}  \ ,  \no \\
 \B^{ab} & = \left( B^{++} T^{++} + B^{--} T^{--} \right)^{ab}  \ ,   \lab{a3}
 \end{align}
 where  $A^{++}$, $A^{--}$,  $B^{++}$,  $B^{--}$  and $A^\pm$  are 6 free  independent  polarization constants. 
 
Doing the same  in the TT gauge \rff{CGtg1},\eqref{CGtg2},\eqref{CGtg3}   leads to 
\begin{align}
\A^{ab}  &= \left(  A^{++}T^{++} + A^{--}T^{--}  \right) ^{ab}   +\left( A^+ (T^+ - \td T^{+}) +A^- (T^- - \td T^{-})
 \right) ^{ab}   \ , \no 
\\
\B^{ab} &= \left( B^{++} T^{++} + B^{--} T^{--} \right) ^{ab}  - 2i \omega \left( A^+ (T^+ +\td  T^{+ })  +  A^-(T^- + \td T^{-}) \right)^{ab} \  .  
\lab{a4} 
\end{align}
While in the   conformal gauge  \rff{a3}  the helicity $\pm1$ states appear  only in the 
  oscillatory A-mode,  in the TT gauge  \rff{a4}  they are  also  present  in  the growing B-mode. Note that it is also possible to make a gauge choice for which the helicity $\pm1$ states appear only in the growing part of the potential. 

\section{Counting  CHS  degrees of freedom}
\label{DOFcount}

In this appendix we demonstrate the counting of on-shell states for the spin-$s$ CHS  field 
 by working directly with (gauge-invariant) field strengths. Let us focus, e.g.,  on the negative helicity sector. 
 The linearised CHS equations of motion in this sector are (cf. \rff{lchs*}) 
\be\label{nh1*}
\partial^{\alpha(s)\dot\alpha(s)}\,\Psi_{\alpha(s)\beta(s)}=0\,,
\ee
with the spinor $\Psi_{\alpha(s)\beta(s)}$ carrying mass dimension $+1$. Our goal is to count linearly independent solutions to these equations; we work in a momentum eigenstate basis 
and in a gauge where solutions take a `helicity lowered' form:
\be  \Psi_{\alpha(s)\beta(s)}=B_{(\alpha(s-1)} \lambda_{\alpha_s)}\lambda_{\beta(s)}\e^{\im\,k\cdot x} \ . \ee
One simple  solution is the standard zero-rest-mass field of helicity $-s$:
\be\label{nhzrm*}
\Psi^{(-s)}_{\alpha(s)\beta(s)}=\kappa_{s}\,\lambda_{\alpha(s)}\,\lambda_{\beta(s)}\,\e^{\im\,k\cdot x}\,,
\ee
where $\kappa_s$ is the mass dimension $1-s$ coupling constant associated with a massless, two-derivative higher spin field. Another solution is provided by
\be\label{nhpo1*}
\Psi^{(-1)}_{\alpha(s)\beta(s)}=a_{(\alpha(s-1)}\,\lambda_{\alpha_s}\,\lambda_{\beta(s)\,)}\,\e^{\im\,k\cdot x}\,,
\ee
where $a^{\alpha(s)}$ is a constant totally symmetric spinor of mass dimension $\frac{1-s}{2}$ which obeys $a_{\alpha(s)}\lambda^{\alpha(s)}\neq 0$. Counting the little group weight in \eqref{nhpo1*} indicates that this solution corresponds to a field of helicity $-1$. 

It is now easy to see that a family of $s$ purely oscillatory solutions is built by taking:
\be\label{nhosc1*}
\Psi^{(-h)}_{\alpha(s)\beta(s)}=\kappa_{h}\,a_{(\alpha(s-h)}\,\lambda_{\alpha(h)}\,\lambda_{\beta(s)\,)}\,\e^{\im\,k\cdot x}\,, \quad \qquad h=1,\ldots,s\,,
\ee
where $\kappa_{h}$ has mass dimension $1-h$ and the constant spinor $a^{\alpha(s-h)}$ has mass dimension $\frac{h-s}{2}$ and obeys $a_{\alpha(s-h)}\lambda^{\alpha(s-h)}\neq 0$. Each such solution is purely oscillatory, and 
$\Psi^{(-h)}$ corresponds to a field of helicity $-h$. These are precisely the $s$ negative helicity scattering states of CHS field of 
 spin $s$.

This leaves us to account for the $\frac{s(s-1)}{2}$ growing states which must make up the remainder of the negative helicity sector. To do this, we first construct the solution of \eqref{nhpo1*} with the highest possible polynomial growth:
\be\label{nhgh1*}
\Psi^{\mathrm{g}(-s)}_{\alpha(s)\beta(s)}=\lambda_{(\beta(s)}\,\lambda_{\alpha_{1}}\,x_{\alpha(s-1)\,)}{}^{\dot\alpha(s-1)}\,\tilde{\beta}_{\dot\alpha(s-1)}\,\e^{\im\,k\cdot x}\,,
\ee
where $\tilde{\beta}_{\dot\alpha(s-1)}$ has mass dimension $\frac{s-1}{2}$ and is constrained so that $\tilde{\beta}_{\dot\alpha(s-1)}\tilde{\lambda}^{\dot\alpha(s-1)}\neq 0$. Counting little group weights tells us that this is a field of helicity $-s$. A further $s-2$ helicity $-s$ growing fields can be constructed by replacing powers of $x$ with insertions of $\lambda$.
 As the number of insertions of $\lambda$ increases, the growth of the field weakens. For example,
  the helicity $-s$ mode with $O(x^{s-l})$ growth is given by:
\be\label{nhgh2*}
\Psi^{\mathrm{g}(-s)}_{\alpha(s)\beta(s)}=\kappa_{l}\lambda_{(\beta(s)}\,\lambda_{\alpha(l)}\,x_{\alpha(s-l)\,)}{}^{\dot\alpha(s-2)}\,\tilde{\beta}_{\dot\alpha(s-l)}\,\e^{\im\,k\cdot x}\,,
\ee
for $\tilde{\beta}_{\dot\alpha{s-l}}$ of mass dimension $\frac{s-l}{2}$. As  $1\leq l\leq s-1$, it is clear that there are in total  $s-1$ 
growing modes of helicity $-s$.

A similar method works for growing modes of helicities $-h$ for $2\leq h<s$. In each case, we simply have to count the number of ways in which the $s$-tuple of spinor indices $\alpha(s)$ can be partitioned among $\lambda_{\alpha}$, $a_{\alpha}$, or $x_{\alpha}{}^{\dot\alpha}\tilde{\beta}_{\dot\alpha}$ insertions. The construction terminates at helicity $-2$, with a single growing field of the form
\be\label{nhgh3*}
\Psi^{\mathrm{g}(-2)}_{\alpha(s)\beta(s)}= \lambda_{(\beta(s)}\,\lambda_{\alpha_1}\,a_{\alpha(s-2)}\,x_{\alpha_s)}{}^{\dot\alpha}\tilde{\beta}_{\dot\alpha}\,\e^{\im\,k\cdot x}\,.
\ee
At each stage, we see that there are $h-1$ distinct forms for the growing mode of helicity $-h$, confirming formula \eqref{CHSdof3} presented above.


\section{Deriving 3-point  amplitudes from twistor space}
\label{PTder}

The 3-point amplitude formulae presented in this paper can be derived in a systematic way from the formulation of CHS theory in twistor space~\cite{Haehnel:2016mlb,Adamo:2016ple}. Here, we review the derivation of the $\overline{\mbox{MHV}}$ 3-point amplitude. This proceeds from the formulation of the self-dual sector of CHS theory in terms of an action functional on twistor space $\PT$, which is an open subset of the three-dimensional complex projective space $\CP^3$, charted with homogeneous coordinates $Z^{A}=(\mu^{\dot\alpha},\lambda_{\alpha})$. The variational data for this twistor action are cohomology classes
\be\label{tdata}
g_{A(s-1)}\in H^{1}(\PT,\cO(-s-3))\,, \qquad f^{A(s-1)}\in H^1(\PT,\cO(s-1))\,,
\ee
which encode the ASD and SD d.o.f. of a spin $s$ CHS field. 

In~\cite{Haehnel:2016mlb,Adamo:2016ple}, it was shown that the SD sector of interacting CHS theory is described in twistor space by the action functional:
\be\label{tact1}
S_{\mathrm{SD}}[g,f]=\int_{\PT}\D^{3}Z\wedge\sum_{|I|=0}^{\infty}g_{A_I}\wedge N^{A_I}\,,
\ee
where $\D^3 Z$ is the holomorphic volume form on $\PT$, and
\be\label{nijen}
N^{A_I}=\dbar f^{A_I}+\sum_{|J|=0}^{|I|}\sum_{|K|=0}^{\infty}\left(\begin{array}{c}
                                                                    |J|+|K| \\
                                                                    |J|
                                                                   \end{array}\right)\,f^{B_{K}(A_J}\wedge \frac{\partial^{|K|}f^{A_{I-J})}}{\partial Z^{B_K}}\,.
\ee
The $\overline{\mbox{MHV}}$ 3-point amplitude is given by extracting the cubic part of this action:
\begin{multline}\label{tmhv1}
\int_{\PT}\D^{3}Z\wedge g_{1\,A_I}\wedge\left[\frac{(|J|+|K|)!}{|K|!\,|J|!}\,f_{2}^{B_K (A_{J}}\wedge\frac{\partial^{|K|}f_{3}^{A_{I-J})}}{\partial Z^{B_K}}\right. \\
\left.+\frac{(|I|-|J|)!}{|K|!\,(|I|-|J|-|K|)!}\,f_{3}^{B_K (A_{I-J-K}}\wedge\frac{\partial^{|K|}f_{3}^{A_{J+K})}}{\partial Z^{B_K}}\right]\,,
\end{multline}
with the multi-indices taking values
\begin{equation*}
 |I|=s_1 -1\,, \quad |J|=s_1-s_3\,, \quad |K|=s_2+s_3-s_1-1\,,
\end{equation*}
in terms of the spins of the external states.

To arrive at a purely momentum space expression, \eqref{tmhv1} is evaluated on the momentum eigenstates:
\be\label{tmeig}
g_{1\,A(s_1-1)}=B_{1\,A(s_1-1)}\,\int \d t_{1}\,t_{1}^{s_1+2}\,\bar{\delta}^{2}(\lambda_{1}-t_{1}\,\lambda)\,\e^{\im\,t_{1}\, [\mu\, 1]}
\ee
\be\label{tmeig2}
f_{i}^{B(s_i-1)}=A_{i}^{B(s_i-1)}\,\int \frac{\d t_{i}}{t_{i}^{s_i}}\,\bar{\delta}^{2}(\lambda_i-t_i\,\lambda)\,\e^{\im\,t_{i}\,[\mu\,i]}\,, \quad i=2,3\,.
\ee
It is easy to see that the action of twistor derivatives on these momentum eigenstates is
\be\label{tmhv2}
\frac{\partial^{|K|}f_{3}^{A_{I-J})}}{\partial Z^{B_K}}=\tilde{C}_{3\,B_K}\,A_{3}^{A_{I-J}}\,\int \frac{\d t_3}{t_{3}^{s_3-|K|}}\,\bar{\delta}^{2}(\lambda_3-t_3\,\lambda)\,\e^{\im\,t_{3}\,[\mu\,3]}\,,
\ee
where $\tilde{C}_{B_K}$ is the momentum space operator \eqref{dopt}. 
This enables us  to express the amplitude as:
\begin{multline}\label{tmhv3}
 \cN^{(\mathbf{s})}\int \frac{\d t_{2}\,\d t_{3}}{t_{2}^{s_2}\,t_{3}^{s_3}} \left[\frac{(s_2-1)!}{(s_1-s_3)!}\,A_{2}^{B_K(A_J}\,\tilde{C}_{3\,B_K}\,B_{1\,A_I}\,A_{3}^{A_{I-J})}\,t_{3}^{|K|}\right. \\
 \left.+(-1)^{s_2+s_3-s_1}\frac{(s_3-1)!}{(s_1-s_2)!}\,A_{3}^{B_K(A_{I-J-K}}\,\tilde{C}_{2\,B_K}\,B_{1\,A_I}\,A_{2}^{A_{J+K})}\,t_{2}^{|K|}\right]\, \delta^{(2)}(\lambda_{2}+t_2\,\lambda_{1}) \\
 \times \delta^{(2)}(\lambda_{3}-t_{3}\,\lambda_1)\,\delta^{(2)}(\tilde{\lambda}_1+t_{2}\,\tilde{\lambda}_2+t_3\,\tilde{\lambda}_3)\,,
\end{multline}
where the overall projective scale has been used to fix $t_1=1$, and the integrals over twistor coordinates $\d^{2}\mu\,\d^{2}\lambda$ have been performed. At this point, the remaining integrals over $\d t_{2}$, $\d t_3$ can be done against the delta functions
$ \delta^{(2)}(\tilde{\lambda}_1+t_{2}\,\tilde{\lambda}_2+t_3\,\tilde{\lambda}_3)\,$
to fix
\begin{equation*}
 t_{2}=\frac{[3\,1]}{[2\,3]}\,, \qquad t_{3}=\frac{[1\,2]}{[2\,3]}\,.
\end{equation*}
The result of this procedure is precisely the formula \eqref{chs3p1} for the $\overline{\mbox{MHV}}$ amplitude in momentum space.


\bibliography{CHS}
\bibliographystyle{JHEP}

\end{document}